\documentclass[12pt]{article}
\usepackage{color}
\usepackage{latexsym}
\usepackage{epsfig,amssymb,euscript, mathrsfs}
\usepackage{amsmath}
\newsavebox{\ns}
\newsavebox{\dbrane}
\newsavebox{\dbshort}

\def\be{\begin{equation}}
\def\ee{\end{equation}}
\def\bea{\begin{eqnarray}}
\def\eea{\end{eqnarray}}

\newcommand{\nn}{\nonumber}

\newcommand\R{\mathbb{R}}
\newcommand\Z{\mathbb{Z}}

\newcommand\diff{\mathrm{d}}
\newcommand{\de}{\partial}

\newcommand{\dd}{\mathrm{d}}
\newcommand{\me}{\mathrm{e}}
\newcommand{\ii}{\mathrm{i}}

\newcommand{\ex}{\mathrm{e}}

\numberwithin{equation}{section}       
\newcommand{\s}{s}
\newcommand\p{\mathcal{P}(p)}
\newcommand\q{\mathcal{Q}(q)}

\newcommand{\bk}{\mathrm{bulk}}
\newcommand{\gr}{\mathrm{grav}}
\newcommand{\ga}{F}  
\newcommand{\bd}{\mathrm{bdry}}
\newcommand{\ct}{\mathrm{ct}}

\newcommand{\free}{\mathcal{F}}
\newcommand{\cov}{\mathscr{D}}
\newcommand{\real}{\mathrm{Re}\, }
\newcommand{\xmin}{x_{\mathrm{min}}}
\newcommand{\xmax}{x_{\mathrm{max}}}
\newcommand{\imag}{\mathrm{Im}\, }

\begin{document}

\begin{titlepage}

\begin{center}

\today

\vskip 2.3 cm 

{\Large \bf The gravity dual of supersymmetric gauge theories}
\vskip .5cm 
{\Large \bf on a squashed three-sphere}

\vskip 2 cm

{Dario Martelli$^1$, Achilleas Passias$^1$ and James Sparks$^2$\\}

\vskip 1cm

$^1$\textit{Department of Mathematics, King's College, London, \\
The Strand, London WC2R 2LS,  United Kingdom\\}

\vskip 0.8cm

$^2$\textit{Mathematical Institute, University of Oxford,\\
24-29 St Giles', Oxford OX1 3LB, United Kingdom\\}

\end{center}

\vskip 2 cm

\begin{abstract}
\noindent We present the gravity dual to a class of three-dimensional $\mathcal{N}=2$ supersymmetric gauge theories on a $U(1)\times U(1)$-invariant squashed 
three-sphere, with a non-trivial background gauge field. 
This is described by a supersymmetric solution of four-dimensional $\mathcal{N}=2$ gauged supergravity with a non-trivial instanton for the graviphoton field. 
The particular gauge theory in  turn determines the lift to a solution of eleven-dimensional supergravity.
We compute the partition function  for a class of Chern-Simons quiver gauge theories on both sides of the duality, 
in the large $N$ limit, finding precise agreement for the functional dependence on the squashing parameter. This constitutes an exact check of the gauge/gravity
correspondence in a non-conformally invariant setting.
 
\end{abstract}

\end{titlepage}

\pagestyle{plain}
\setcounter{page}{1}
\newcounter{bean}
\baselineskip18pt

\tableofcontents

\section{Introduction and summary}

There has been considerable interest recently in studying supersymmetric gauge theories  on curved manifolds. Quite generally, supersymmetric 
field theories on compact curved backgrounds are particularly amenable to localization techniques, leading to vast simplifications in the  exact computation of
partition functions and other observables in strongly coupled field theories. The partition functions of $\mathcal{N}=2$ gauge theories on $S^4$ and certain 
Wilson loops were computed  in \cite{Pestun:2007rz}. Using similar techniques, the partition functions of three-dimensional $\mathcal{N}=3$ supersymmetric gauge 
theories on a round $S^3$ were first computed in \cite{Kapustin:2009kz},  and subsequently generalized to ${\cal N}=2$ gauge theories in \cite{Jafferis:2010un,Hama:2010av}. 
One can also consider curved manifolds other than round spheres.  For example, the superconformal indices of four-dimensional and three-dimensional field theories 
may be computed  by putting the theories  on $S^1\times S^3$  \cite{Romelsberger:2005eg,Dolan:2008qi,Gadde:2010en} and $S^1\times S^2$  
\cite{Imamura:2011uj,Cheon:2011th,Gadde:2011ia, Imamura:2011uw}, respectively. 

A more systematic analysis of the possible curved manifolds on which 
one can construct supersymmetric theories has been initiated in  \cite{Festuccia:2011ws}.  
One particularly interesting possibility
is that of deformed three-spheres, often referred to as \emph{squashed} three-spheres.
The partition functions of three-dimensional $\mathcal{N}=2$ supersymmetric gauge theories on 
different squashed spheres were computed by Hama, Hosomichi and Lee (HHL) in \cite{Hama:2011ea}. 
An interesting deformation in \cite{Hama:2011ea} preserves a $U(1)\times U(1)$ isometry, 
and in the partition function leads to the appearance of the double sine function $s_b(z)$ \cite{barnes}, 
also referred to as the quantum dilogarithm function. This special function plays an important role in various contexts. 
For example,  the double sine function and the $U(1)^2$-squashed sphere (which is a three-dimensional ellipsoid) are important 
ingredients in the AGT correspondence \cite{Alday:2009aq} and its 3d/3d version \cite{Hosomichi:2010vh, Dimofte:2011ju}.
The matching of  partition functions also allows one to perform non-trivial tests of  conjectured dualities between pairs of three-dimensional 
field theories \cite{Kapustin:2010mh,Kapustin:2011gh,Willett:2011gp,Dolan:2011rp,Jafferis:2011ns,Krattenthaler:2011da,Benvenuti:2011ga,Nishioka:2011dq, Benini:2011mf}.

Knowledge of the exact partition functions of three-dimensional Chern-Simons matter theories has also been key for some of the recent non-trivial tests of the 
AdS$_4$/CFT$_3$ correspondence. In \cite{Drukker:2010nc} the free energy of the ABJM matrix model arising from localization on the round three-sphere was 
matched to the dual holographic free 
energy in the large $N$ limit, in particular reproducing the famous $N^{3/2}$ gravity prediction from a purely field theoretic computation. This matching 
was extended in \cite{Herzog:2010hf,Santamaria:2010dm} to examples with ${\cal N}=3$ supersymmetry,
 and then subsequently to a large class of ${\cal N}=2$ models in  \cite{Martelli:2011qj, Cheon:2011vi, Jafferis:2011zi}. 
It is then natural to attempt to construct the gravity dual of $\mathcal{N}=2$ Chern-Simons quiver theories on the $U(1)^2$-squashed three-sphere, 
and to compare the holographic free energy with the large $N$ behaviour of the field theoretic free energy obtained from the HHL matrix integral.  
In this paper we will address this problem for a large class of  
$\mathcal{N}=2$ Chern-Simons theories, finding exactly the same non-trivial 
dependence on the deformation parameter on the two sides.   

\subsubsection*{Preview}

In \cite{Hama:2011ea}, HHL have shown that rigid ${\cal N }=2$ supersymmetric Chern-Simons gauge theories 
can be put on a $U(1)^2$-invariant squashed three-sphere, by appropriately modifying the Lagrangian and supersymmetry transformations.  
In particular, the metric used in \cite{Hama:2011ea} is, up to an irrelevant overall constant factor, given by
\bea\label{HHLintro}
\diff s^2_{3} &=& f^2(\theta) {\diff \theta^2}+\cos^2\theta\diff\varphi_1^2 + \frac{1}{b^4}\sin^2\theta\diff\varphi_2^2~,
\eea
where the squashing parameter is $b^2=\ell/\tilde{\ell}$, 
and the function $f^2(\theta)$ will be specified below.  
The spinor parameter $\chi$ entering the supersymmetry transformations obeys the modified 
Killing spinor equation
\bea\label{3dKSEintro}
(\nabla_\alpha^{(3)} - \ii  A_\alpha^{(3)})\chi - \frac{\ii}{2f(\theta)}\gamma_\alpha \chi &=& 0~,
\eea
where $\nabla_\alpha^{(3)}$,   $\alpha=1,2,3$, is the spinor covariant derivative constructed from the metric (\ref{HHLintro}), 
$\gamma_\alpha$  generate $\mathrm{Cliff}(3,0)$,
and 
\bea\label{boundarygaugeintro}
A^{(3)} &=&  \frac{1}{2f(\theta)}\left(\diff\varphi_1 -\frac{1}{b^2}\diff\varphi_2\right)
\eea
is a background gauge field.\footnote{Written here up to an irrelevant gauge transformation, $A^{(3)} = A^{(3)}_\mathrm{HHL} +\tfrac{1}{2}(\dd \varphi_1 -\dd \varphi_2)$.}
In  \cite{Hama:2011ea} the squashed three-sphere (\ref{HHLintro}) arises as the metric induced on the hypersurface
\bea\label{hypersurfaceintro}
r_1^2 + b^4 r_2^2 & = & r^2
\eea
in flat $\R^4=\R^2\oplus \R^2$ with metric
\bea
\diff s^2_{\R^4} &=& \diff r_1^2 + \diff r_2^2+ r_1^2 \dd\varphi_1^2 + r_2^2\diff\varphi_2^2~. 
\eea
Here one can take $r>0$ to be any constant, although the metric in (\ref{HHLintro}) is normalized so that $r=1$.
This leads to the particular function $f^2(\theta) = \sin^2\theta+\tfrac{1}{b^4}\cos^2\theta$, 
and by definition (\ref{HHLintro}) is then the metric on an \emph{ellipsoid}. 
However, notice that 
(\ref{HHLintro}) is a non-singular metric on $S^3$ for \emph{any} strictly positive (or negative) function $f(\theta)$ that approaches sufficiently smoothly 
$|f(\theta)|\to 1/b^2$ as $\theta \to 0$ and  $|f(\theta)|\to 1$ as $\theta \to \tfrac{\pi}{2}$.  This observation will 
be important in what follows.

The main result of 
\cite{Hama:2011ea} is that the partition function of a supersymmetric gauge theory in the background of 
(\ref{HHLintro})  \emph{and} the gauge field (\ref{boundarygaugeintro}) can be computed using localization techniques, and reduces to 
a matrix integral generalizing that of \cite{Jafferis:2010un,Hama:2010av}. As we will describe in more detail in 
section \ref{localsection}, for 
a $U(N)^G$ Chern-Simons quiver gauge theory with Chern-Simons levels $k_I$, $I=1,\ldots, G$, the partition function  reads
\bea
Z_b & = & \frac{1}{N!^G}\int\left(\prod_{I=1}^G\prod_{i=1}^N\frac{\diff\lambda_i^I}{2\pi}\right)\exp\left[-F_b\left(\lambda_i^I\right)\right]~,
\label{partfun}
\eea
where
\bea\label{Fintro}
F_b\left(\lambda_i^I\right) &=&- \frac{\ii}{b^2}\sum_{I=1}^G\frac{k_I}{4\pi}\sum_{i=1}^N (\lambda_i^I)^2 - \sum_{I=1}^G \sum_{i< j} \Bigg[
\log \left(2\sinh \frac{\lambda_i^I-\lambda_j^I}{2}\right) \\
&& +\log \left(2\sinh \frac{\lambda_i^I-\lambda_j^I}{2b^2}\right)\Bigg]
 - \sum_{I\rightarrow J}\sum_{i,j=1}^N s_b\left[\frac{\ii Q}{2}(1-\Delta_{I,J})-\frac{(\lambda_i^I-\lambda_j^J)}{2\pi b}\right]~.\nn
\eea
The first term in (\ref{Fintro}) comes from the classical Chern-Simons action, while the second term is a one-loop 
contribution from the gauge field multiplet. The final one-loop term in (\ref{Fintro}) contains a sum over 
bifundamental fields in the fundamental of the $I$th gauge group factor and anti-fundamental 
of the $J$th, of R-charge $\Delta_{I,J}$, and we have defined
\bea\label{Qdef}
Q &\equiv & b+ \frac{1}{b}~.
\eea
Importantly, it turns out that this result \emph{does not} depend on the details of the function $f(\theta)$.
In section \ref{localsection} we will review the localization calculation of \cite{Hama:2011ea}, emphasizing its
 independence of the precise choice of $f(\theta)$.

In this paper we will present the gravity dual to the set-up described above. In particular, we will discuss a 1/4 supersymmetric solution of $d=4$, 
${\cal N}=2$ gauged supergravity (Einstein-Maxwell theory) that asymptotically approaches the metric (\ref{HHLintro}) and gauge field 
(\ref{boundarygaugeintro}), albeit with a function $f(\theta)$ that is different from that used in \cite{Hama:2011ea}. 
Indeed, while the HHL ellipsoid metric arises from the hypersurface (\ref{hypersurfaceintro}) in \emph{flat space}, 
instead our boundary three-metric arises from the \emph{same} hypersurface equation (\ref{hypersurfaceintro}), but now in 
\emph{hyperbolic space} $\mathbb{H}^4$ (Euclidean AdS$_4$) with metric
\bea\label{hyperbolicmetric}
\diff s^2_{\mathbb{H}^4} \, \, =\, \, \frac{1}{r_1^2+r_2^2+1}\left[\diff r_1^2  
+ \diff r_2^2 + (r_2\diff r_1-r_1\diff r_2)^2\right]+ r_1^2\diff\varphi_1^2+ r_2^2\diff\varphi_2^2~.
\eea
More precisely, our three-metric arises from the limit $r\rightarrow \infty$ in (\ref{hypersurfaceintro}),  which  leads to 
the particular function $f^2(\theta)=1/(b^4\cos^2\theta+\sin^2\theta)$ in (\ref{HHLintro}).
We may therefore refer to our particular squashed $S^3$ as a \emph{hyperbolic ellipsoid}. 
Of course, by construction it arises as the conformal boundary of Euclidean AdS$_4$ (\ref{hyperbolicmetric}), and thus 
unlike the HHL metric in reference \cite{Hama:2011ea} our squashed $S^3$ metric is conformal to the round metric on $S^3$. 
However, 
it will also be important to turn on an appropriate $U(1)$ instanton. This then uplifts to a supersymmetric solution of eleven-dimensional 
supergravity of the form $M_4\times Y_7$, where $Y_7$ is any Sasaki-Einstein
seven-manifold, and the product is twisted \cite{Gauntlett:2007ma, Gauntlett:2009zw}. (It also  uplifts \cite{Gauntlett:2007ma} to a solution 
of the twisted, warped form $M_4\times_w N_7$, corresponding to M5-branes wrapping SLag three-cycles, although we will not use this in this paper.)
Moreover,  the Killing spinor for the four-dimensional supergravity solution, restricted to the boundary, 
precisely solves  the equation (\ref{3dKSEintro}). 

 We will then  compute the holographic free 
energy for this supergravity solution and compare it with the large $N$ limit of the free energy of a large class of ${\cal N}=2$ Chern-Simons quiver theories, obtained from (\ref{partfun}).  
We will find exact agreement, and in particular  in both cases we will show that the free energy satisfies
\bea
\free_b & = & \frac{Q^2}{4} \free_{b=1}~,
\eea
where $Q$ is given by (\ref{Qdef}). The  $U(1)$ gauge field instanton breaks explicitly the symmetries of the conformal group 
$SO(2,3)$. Therefore, this constitutes an exact check of the gauge/gravity correspondence in a non-conformally invariant setting.

The rest of this paper is organized as follows. In section \ref{gravityside} we discuss the gravity side: we present the solution and compute its
holographic free energy. In section \ref{fieldtheoryside} we discuss the field theory side: we review the computation of the partition function and extract
the large  $N$ limit of the free energy. Section \ref{discussionsex} concludes. In appendix 
\ref{PDappendix} we discuss the solution in the more general context of Plebanski-Demianski solutions to Einstein-Maxwell theory. Appendices \ref{Killappendix} 
and \ref{vectorapp} contain some technical computational details for the Killing spinor and one-loop vector multiplet 
contribution to the partition function, respectively.

\section{The gravity side}

\label{gravityside}

\subsection{$d=4$, $\mathcal{N}=2$ gauged supergravity and $d=11$ uplift}\label{gaugedsec}

Our starting point is the action for the bosonic sector of $d=4$, $\mathcal{N}=2$ gauged supergravity \cite{Freedman:1976aw}
\bea\label{4dSUGRA}
S &=& -\frac{1}{16\pi G_4}\int \diff^4x\sqrt{\det g_{\mu\nu}}\left(R + 6g^2 - F^2\right)~.
\eea 
Here $R$ denotes the Ricci scalar of the four-dimensional metric $g_{\mu\nu}$, and the cosmological constant 
is given by $\Lambda=-3g^2$. The graviphoton is an Abelian gauge field $A$ with field strength 
$F=\diff A$, and we have denoted $F_{\mu\nu}F^{\mu\nu}=F^2$. 
We will mainly be working in Euclidean signature, and have denoted the four-dimensional
Newton constant by $G_4$.

A solution to the equations of motion derived from (\ref{4dSUGRA}), namely
\bea\label{EOM}
R_{\mu\nu} &=& -3g^2g_{\mu\nu} +2\left(F_\mu^{\ \rho}F_{\nu\rho}-\tfrac{1}{4}F^2 g_{\mu\nu}\right)~,\nonumber\\
\diff *_4F &=&0~,
\eea
 is supersymmetric if there is a non-trivial Dirac spinor $\epsilon$
satisfying the Killing spinor equation
\bea\label{KSE}
\left[ \nabla_\mu + \tfrac{1}{2} g \Gamma_\mu - \ii g A_\mu + \tfrac{\ii}{4} F_{\nu\rho} \Gamma^{\nu\rho} \Gamma_\mu \right] \epsilon &=& 0~.
\eea
Here $\Gamma_\mu$, $\mu=1,2,3,4$, generate the Clifford algebra $\mathrm{Cliff}(4,0)$, so 
$\{\Gamma_\mu,\Gamma_\nu\}=2g_{\mu\nu}$.

As shown in \cite{Gauntlett:2007ma, Gauntlett:2009zw}, any such solution to $d=4$, $\mathcal{N}=2$ 
gauged supergravity uplifts to a supersymmetric solution of eleven-dimensional 
supergravity. More precisely, given any Sasaki-Einstein seven-manifold $Y_7$ with contact one-form $\eta$, transverse K\"ahler-Einstein metric $\diff s^2_T$, 
and with the seven-dimensional metric normalized so that $R_{ij}=6g_{ij}$, we write
\bea\label{uplift}
\diff s^2_{11} &=& R^2\left(\tfrac{1}{4}\diff s^2_4 + \left(\eta + \tfrac{1}{2}A\right)^2+\diff s^2_T\right)~,\nonumber\\
G &=& R^3\left(\tfrac{3}{8}\mathrm{vol}_4 - \tfrac{1}{4}*_4 F \wedge \diff \eta\right)~.
\eea
Here $\diff s^2_4$ is the four-dimensional gauged supergravity metric, with volume form $\mathrm{vol}_4$, 
and we have set $g=1$. The effective AdS$_4$ radius $R$ is then determined by the quantization of 
the four-form flux $G$ via
\bea
N &=& \frac{1}{(2\pi \ell_p)^6}\int_{Y_7} *_{11} G~,
\eea
where $\ell_p$ is the eleven-dimensional Planck length, which leads to
\bea
R^6 &=& \frac{(2\pi\ell_p)^6N}{6\mathrm{Vol}(Y_7)}~.
\eea
The effective four-dimensional Newton constant is then
\bea\label{Newton}
\frac{1}{16\pi G_4} &=& N^{3/2}\sqrt{\frac{\pi^2}{32\cdot 27\, \mathrm{Vol}(Y_7)}}~.
\eea

In fact it was more generally conjectured in \cite{Gauntlett:2007ma} that 
given any  $\mathcal{N}=2$  warped AdS$_4\times Y_7$ solution of eleven-dimensional supergravity there is a 
consistent Kaluza-Klein truncation on $Y_7$ to the above $d=4$, $\mathcal{N}=2$ gauged supergravity theory. 
Properties of such general solutions have recently been investigated in \cite{Gabella:2011sg}, and we 
expect the contact structure discussed there to play an important role in this truncation. In particular, 
it was shown in \cite{Gabella:2011sg} that (\ref{Newton}) remains true in this more general setting, 
provided one replaces the Riemannian volume by the contact volume.

\subsection{Supergravity solution}
\label{sugrasolun}

We will be interested in the following supersymmetric solution to the above $d=4$, $\mathcal{N}=2$ gauged supergravity theory:
\bea\label{soln}
\diff s^2_4 &=& f_1^2(x,y)\diff x^2 + f_2^2(x,y)\diff y^2 + \frac{(\diff\Psi+y^2\diff\Phi)^2}{f_1^2(x,y)}+ \frac{(\diff\Psi+x^2\diff\Phi)^2}{f_2^2(x,y)}~,\nonumber\\
A &=& g(\s^2-1) \frac{\diff\Psi - xy\diff\Phi}{2(y+x)}~,
\eea
where we have defined the functions
\bea
f_1^2(x,y) \  \equiv \  \frac{y^2-x^2}{g^2(x^2-1)(\s^2-x^2)} ~,~~~ f_2^2(x,y) \  \equiv  \ \frac{y^2-x^2}{g^2(y^2-1)(y^2-\s^2)}~.
\eea
This arises as a special case of the class of Plebanski-Demianski solutions to Einstein-Maxwell theory \cite{PD}, 
whose supersymmetry was investigated (in Lorentzian signature) in  \cite{AlonsoAlberca:2000cs}. 
However, for our purposes it will be crucial to obtain the explicit form of the Killing spinor of this solution, in the context of ${\cal N}=2$ gauged 
supergravity, and as far as we are aware this analysis is new.  

The solution depends on one parameter $\s$, which will take the values $\s\in [1,\infty)$. 
In fact, as anticipated in the introduction, the metric in (\ref{soln}) is locally just the (Euclidean) AdS$_4$ metric, for any value of $\s$, as is 
easily verified by 
checking that the Riemann curvature tensor obeys $R_{\mu\nu\rho\sigma} 
= -g^2(g_{\mu\rho}g_{\nu\sigma} - g_{\mu\sigma}g_{\nu\rho})$. The unusual 
coordinate system in (\ref{soln}) is inherited from its origin as a Plebanski-Demianski 
solution \cite{PD}, as discussed further in appendix \ref{PDappendix}. We shall also make use of the following coordinates:
\bea
\frac{x^2-1}{\s^2-1}\  \equiv \ \cos^2\theta~, 
\qquad \Psi \ \equiv  \ \frac{\s\varphi_2-\varphi_1}{g^2(\s^2-1)}~,  
\qquad \Phi & \equiv & \frac{\s\varphi_1-\varphi_2}{\s g^2(\s^2-1)}~.
\eea
Introducing also
\bea
h^2(\theta)&\equiv & \s^2\cos^2\theta+\sin^2\theta~,
\eea
the four-dimensional  metric in (\ref{soln}) becomes
\bea\label{thetametric}
\diff s^2_4 &= & \frac{y^2-h^2(\theta)}{g^2(y^2-1)(y^2-\s^2)}\diff y^2 +\frac{y^2-h^2(\theta)}{g^2h^2(\theta)}\diff \theta^2 +
 \frac{y^2-1}{g^2}\cos^2\theta\diff\varphi_1^2  \nonumber\\&&
+ \frac{y^2-\s^2}{\s^2 g^2}\sin^2\theta\dd\varphi_2^2~.
\eea
Here the ranges of the coordinates are $y\in [\s,\infty)$, $\theta\in[0,\tfrac{\pi}{2}]$, while $\varphi_1$ and $\varphi_2$ are periodic 
with period $2\pi$. In particular this implies that $x\in [1,\s]$ (when $s>1$). This will be discussed further in section \ref{globalsec}. 

Introducing the orthonormal frame
\bea\label{frame}
e^1 \ = \ \frac{\diff\Psi + y^2 \diff\Phi}{f_1} ~, ~~~~~~~ e^3 \ = \ f_1 \diff x~,\nonumber\\
e^2 \ = \ \frac{\diff\Psi + x^2 \diff\Phi}{f_2}  ~, ~~~~~~~ e^4 \ = \ f_2 \diff y~,
\eea
the gauge field in (\ref{soln}) has field strength
\bea\label{fieldstrength}
F &=& \diff A \ = \ \frac{g(\s^2-1)}{2(y+x)^2}(e^{13}+e^{24})~.
\eea
In particular, we see that $F$ is anti-self-dual, $*_4F=-F$, and hence that $A$ is an instanton.   We will see in section \ref{holosec} that the action is  indeed finite.

Notice that when $s=1$ the gauge field strength is zero, and moreover the metric as presented in (\ref{thetametric}) is 
more obviously the metric on Euclidean AdS$_4$, since in this case $h^2(\theta)\equiv 1$ and 
\bea\label{AdS4metric}
\diff s^2_4\mid_{\s=1}&=& \frac{\diff y^2}{g^2(y^2-1)}+\frac{y^2-1}{g^2}(\diff\theta^2 + \cos^2\theta\diff\varphi_1^2+\sin^2\theta\diff\varphi_2^2)~.
\eea
This describes Euclidean AdS$_4$ as a hyperbolic ball with boundary conformal to the round metric on $S^3$, the latter appearing in the round brackets.
When $s>1$ the metric (\ref{thetametric}) continues to be a smooth complete metric on AdS$_4$, but with $y$ being a different 
choice of radial coordinate to that in (\ref{AdS4metric}), as we shall see in the subsection below. Of course, for $s>1$ we are also turning on a non-trivial 
instanton in the graviphoton field, via (\ref{fieldstrength}).

\subsection{Global structure}
\label{globalsec}

At large values of $y$ the metric (\ref{thetametric}) is
\bea\label{asymptoticmetric}
\diff s^2_4 &=& \frac{\diff y^2}{g^2y^2}\left[1+O(\tfrac{1}{y^2})\right]+\frac{y^2}{g^2}\left[\diff s^2_3 +O(\tfrac{1}{y^2})\right]~,
\eea
where
\bea\label{squashedS3}
\diff s^2_3 &=& \frac{\diff \theta^2}{s^2\cos^2\theta+\sin^2\theta}+\cos^2\theta\diff\varphi_1^2 + \frac{1}{\s^2}\sin^2\theta\diff\varphi_2^2~.
\eea
Thus our Euclidean AdS$_4$ metric has as conformal boundary at $y=\infty$ the metric of a squashed $S^3$ of the 
form (\ref{HHLintro}), where $\s=b^2 \in [1,\infty)$ is the squashing 
parameter and $f^2(\theta)=1/h^2(\theta)=1/(s^2\cos^2\theta+\sin^2\theta)$. In particular, for $s=1$ we recover the round metric on $S^3$, as already noted. 

Returning to the full four-dimensional metric (\ref{thetametric}), it is immediate to see that
we obtain a smooth induced metric on $S^3$ at \emph{any} value of $y\in (\s,\infty)$, and that this hence 
defines a smooth, but incomplete, metric on $\R_{>0}\times S^3$, where $y-\s$ is a coordinate on $\R_{>0}$. 
It thus remains to examine what happens as $y$ tends to $\s$ from above.  Although one can examine this directly in the above coordinates, it is easier to see what is going on globally by 
changing coordinates again:
\bea\label{coordchange}
r_1^2 & \equiv & (y^2-1)\cos^2\theta~,\nonumber\\
r_2^2 & \equiv & \frac{1}{s^2}(y^2-s^2)\sin^2\theta~.
\eea
The metric (\ref{thetametric}) (multiplied by $g^2$) then becomes the Euclidean AdS$_4$ metric (\ref{hyperbolicmetric}) 
presented in the introduction. 
The parameter $\s$ has disappeared, and we directly see the local equivalence to the Euclidean AdS$_4$ metric for all $\s\in [1,\infty)$. 
Notice that for $y$ large, (\ref{coordchange}) gives $y^2 \simeq r_1^2+s^2 r_2^2$, 
as claimed in the introduction. Also notice that for $s>1$, $y=s$ is simply the coordinate singularity $r_2=0$. 
The ``centre'' of AdS$_4$ in the coordinates (\ref{hyperbolicmetric}) is $\{r_1=r_2=0\}$, which 
is $\{y=s, \theta=\frac{\pi}{2}\}$.

It follows from this discussion that our metric (\ref{thetametric}) is simply the metric on the usual Euclidean AdS four-ball, but 
with a non-standard choice of radial coordinate $y$. In particular, this means that the conformal class of the induced metric 
on $y=\infty$ is that of a squashed $S^3$, with metric given by (\ref{squashedS3}). In other words, we have a one-parameter family of ``faces of AdS'', to 
use the  phrase coined in \cite{Emparan:1999pm}, given by choosing a radial coordinate that depends on $s$. 

Of course, it will also be important for our application
that the instanton in (\ref{soln}) depends on $s$. Notice that the field strength $F$ in (\ref{soln}) is a  non-singular globally defined  two-form 
on our Euclidean AdS ball. Indeed, a short computation gives
\bea
F &=& \frac{s^2-1}{2g(y+h(\theta))^2}\Big[ \left(\cos^2\theta \diff\varphi_1 + s^{-1}\sin^2\theta\diff\varphi_2\right)\wedge \diff y 
\nonumber\\
&& - \left[(y^2-1)\diff\varphi_1- s^{-1} (y^2-s^2)\diff\varphi_2\right]\sin\theta\cos\theta\wedge \frac{\diff \theta}{h(\theta)} \, \Big]~.
\eea
In particular, notice that $y+x$ is nowhere zero, since 
both $y$ and $x=h(\theta)$ are strictly positive. On the other hand, 
there is a \emph{self-dual} instanton for the same metric given by
\bea
F_{\mathrm{SD}} &=& \frac{g(\s^2-1)}{2(y-x)^2}(e^{13}-e^{24})~.
\eea
Compare this with (\ref{fieldstrength}). However, since $y=x$ on the locus $\{y=s, x=s\}$ (or equivalently 
$\{y=s,\theta=0\}$), this self-dual instanton is singular on this locus. 

\subsection{Supersymmetry}\label{susysec}

In this subsection we discuss the supersymmetry of the solution. 
That the solution is supersymmetric is perhaps not surprising, given that we simply have an instanton on AdS$_4$ space. 
However, instantons work a little differently in AdS than on Ricci-flat manifolds, due to the cosmological constant, 
and the precise form of the Killing spinor on our background (particularly its asymptotic expansion) will be important 
in section \ref{fieldtheoryside}. It is perhaps worth noting that the Killing spinor $\epsilon$ 
solving (\ref{KSE}) is \emph{not} (and even {\it a priori} could not be) one of the usual Killing spinors of AdS$_4$.

It will be convenient to choose the following representation of $\mathrm{Cliff}(4,0)$:
\bea\label{gammas}
\hat{\Gamma}_a & =  & \begin{pmatrix} 0 & \sigma_a \\ \sigma_a & 0 \end{pmatrix} ~, 
\qquad \qquad \hat{\Gamma}_4 \ =  \ \begin{pmatrix} 0 & \ii \mathbb{I}_2 \\ -\ii \mathbb{I}_2 & 0 \end{pmatrix}~,
\eea
where $\sigma_{a}$, $a=1,2,3$, denote the Pauli matrices, and hats denote tangent space quantities. 
Thus $\{\hat{\Gamma}_m,\hat{\Gamma}_n\}=2\delta_{mn}$. In particular then 
$\hat{\Gamma}_5\equiv \hat{\Gamma}_1\hat{\Gamma}_2\hat{\Gamma}_3\hat{\Gamma}_4$ 
is given by
\bea
\hat{\Gamma}_5 & =  & \begin{pmatrix}\  \mathbb{I}_2 &0 \\ 0 & -\mathbb{I}_2 \ \end{pmatrix} ~, 
\eea
and we may decompose the Dirac spinor $\epsilon$ into negative and positive chirality 
parts as 
\bea
\epsilon &=& \left(\begin{array}{c}\epsilon^+ \\ \epsilon^-\end{array}\right)~.
\eea
One can then substitute into the Killing spinor equation (\ref{KSE}) using the orthonormal 
frame (\ref{frame}). In fact an immediate consequence of the integrability condition 
for (\ref{KSE}) is the relation
\bea
2g F_{\mu\nu} \epsilon^+ &=& (\nabla^\rho F_{\mu\nu})\gamma_\rho\epsilon^-~,
\eea
where we have defined $\hat{\gamma}_a = \sigma_a$, $a=1,2,3$, $\hat{\gamma}_4= \ii \mathbb{I}_2$, 
and made use of the Bianchi identity for $F$ and that the metric is  AdS$_4$ . In the present case this may be rewritten
\bea\label{algebraic}
\epsilon^+ &=& -\frac{1}{g(y+x)}\left(\frac{\ii}{f_2}\mathbb{I}_2 + \frac{1}{f_1}\sigma_3\right)\epsilon^-~,
\eea
allowing us to algebraically eliminate $\epsilon^+$ in terms of $\epsilon^-$. It is then straightforward, but 
somewhat tedious, to verify that
\bea\label{epplus}
\epsilon^- &=& \sqrt{y+x}\left(\begin{array}{c} \lambda(x,y)\\ \ii \lambda^*(x,y)\end{array}\right)~,
\eea
where 
\bea
\lambda(x,y) &\equiv & \left(\frac{\sqrt{(\s^2-x^2)(y^2-1)}-\ii \sqrt{(x^2-1)(y^2-\s^2)}}{\sqrt{(\s^2-x^2)(y^2-1)}+\ii \sqrt{(x^2-1)(y^2-\s^2)}}\right)^{1/2}~,
\eea
is the only solution to the Killing spinor equation (\ref{KSE}), up to a constant of proportionality. Note in particular that 
the Killing spinor is, in the gauge where $A$ takes the form in (\ref{soln}), independent of the angular coordinates 
$\Psi$ and $\Phi$ (or equivalently $\varphi_1$ and $\varphi_2$).

The solution thus preserves $\mathcal{N}=1$ supersymmetry, in the sense that it admits a single 
Dirac spinor $\epsilon$ solving (\ref{KSE}). However, it will be important later that 
the charge conjugate spinor $\epsilon^c \equiv B\epsilon^*$ also satisfies the 
Killing spinor equation (\ref{KSE}) but with $A$ replaced by $-A$. Here $B$ is the charge conjugation matrix
\bea
B &\equiv & \left(\begin{array}{cc}\varepsilon & 0 \\ 0 & -\varepsilon\end{array}\right)~, \qquad \quad \varepsilon \ \equiv \ \left(\begin{array}{cc} 0 & -1 \\ 1 & 0 \end{array}\right)~, 
\eea
which satisfies the defining properties
\bea
B^{-1} {\Gamma}_\mu B &=& \Gamma_\mu^*~, \qquad \quad BB^* \ = -\mathbb{I}_4~.
\eea
Thus it is clear that provided $\epsilon$ satisfies (\ref{KSE}), then $\epsilon^c$ satisfies
\bea\label{KSEbar}
\left[ \nabla_\mu + \tfrac{1}{2} g \Gamma_\mu +\ii g A_\mu - \tfrac{\ii}{4} F_{\nu\rho} \Gamma^{\nu\rho} \Gamma_\mu \right] \epsilon^c &=& 0~.
\eea

At large $y$ it is straightforward to calculate the asymptotic expansion of the Killing spinor (\ref{epplus}). Still working 
in the frame (\ref{frame}) one finds
\bea
\epsilon &=& \left(\begin{array}{c} - y^{1/2}\left[1-\frac{h(\theta)}{2y}+O(\tfrac{1}{y^2})\right]\ii\chi\\ 
y^{1/2}\left[1+\frac{h(\theta)}{2y}+O(\tfrac{1}{y^2})\right]\chi\end{array}\right)~,
\eea
where
\bea\label{originalframe}
\chi &=& \left(\begin{array}{c}\ii \ex^{\ii\theta}\\ \ex^{-\ii\theta}\end{array}\right)~.
\eea
The latter defines a spinor on the squashed three-sphere conformal boundary, and one finds that 
$\chi$ satisfies the following Killing spinor equation on the squashed sphere (\ref{squashedS3})
\bea\label{3dKSE}
(\nabla_\alpha^{(3)} - \ii g A_\alpha^{(3)})\chi + \frac{\ii h(\theta)}{2}\gamma_\alpha \chi &=& 0~.
\eea
Here $\gamma_\alpha$ generate $\mathrm{Cliff}(3,0)$, $\alpha=1,2,3$, while $A^{(3)}$ denotes 
the asymptotic value of the gauge field in (\ref{soln}), namely
\bea\label{boundarygauge}
A^{(3)} &=& - g(s^2-1) \frac{x}{2}\diff\Phi~,\nonumber\\
& =& -\frac{h(\theta)}{2g}\left(\diff\varphi_1 -\frac{1}{s}\diff\varphi_2\right)~.
\eea
Notice that (\ref{3dKSE}), (\ref{boundarygauge}) are precisely of the form (\ref{3dKSEintro}), (\ref{boundarygaugeintro}) 
in the introduction, on identifying $f(\theta)=-1/h(\theta)$.

The Killing spinor in (\ref{originalframe}) is in the somewhat unusual frame
\bea\label{3dframe}
\hat{e}^1 &=& \frac{1}{s}\cos\theta\sin\theta (s\dd\varphi_1 - \dd\varphi_2)~, \quad 
\hat{e}^2 \, = \,  \frac{1}{s}\left[\dd \varphi_2 +\cos^2\theta(s\dd\varphi_1 - \dd\varphi_2)\right]~,\nn\\
\hat{e}^3 &=& -\frac{\dd\theta}{h(\theta)}~,
\eea
inherited from (\ref{frame}).
It is clearly more natural to define the following orthonormal frame for the squashed $S^3$ 
\bea\label{squashedframe}
\check{e}^1 &=& \cos\theta\, \diff\varphi_1~,\qquad \check{e}^2 \ =\ \frac{1}{s}\sin\theta\, \diff\varphi_2~, \qquad \check{e}^3 \ = \ -\frac{\diff\theta}{h(\theta)}~.
\eea
In this frame, and with $\gamma_\alpha = \check{e}_\alpha^a \sigma_a$, one finds that the solution to (\ref{3dKSE}) is
\bea\label{KJKS}
\chi &=& \ex^{\ii\pi/4}\left(\begin{array}{c}\ex^{\ii\theta/2}\\ \ex^{-\ii\theta/2}\end{array}\right)~.
\eea
This is of course related to (\ref{originalframe}) by a $U(1)\subset SU(2)$ rotation that covers the $SO(2)\subset SO(3)$ 
rotation relating the frame (\ref{squashedframe}) to the corresponding frame given by (\ref{3dframe}). 
Notice that, in this frame, the Killing spinor (\ref{KJKS}) is independent of the squashing parameter $s$, and 
is identical (up to an irrelevant proportionality constant and the gauge transformation in footnote 1) to the Killing spinor $\bar\epsilon$ in 
section 2 of \cite{Hama:2011ea}. The Killing spinor $\epsilon$ of \cite{Hama:2011ea} coincides with the charge conjugate $\chi^c$ 
which satisfies the same Killing spinor 
equation but with $A^{(3)}$ replaced by $-A^{(3)}$.

\subsection{AdS$_4$ with round $S^3$ boundary}\label{roundsec}

As we argued in sections \ref{sugrasolun} and \ref{globalsec},
 our four-dimensional metric is in fact globally Euclidean anti de Sitter space. In this section we elaborate on this point, presenting the 
solution in more standard coordinates, and discussing the induced background gauge field and Killing spinors on 
the boundary.\footnote{This section was added in version 2 of the preprint in March 2012.
 We wish to thank Jerome Gauntlett and David Tong for discussions that prompted us to add this section.}

A more standard coordinate system on AdS$_4$ is obtained by defining
\bea
q^2 & \equiv & r_1^2+ r_2^2~, \qquad \cos^2\psi \ \equiv \ \frac{r_1^2}{r_1^2+r_2^2}~,
\label{newcoc}
\eea
where $r_1, r_2$ were defined in (\ref{coordchange}). In these coordinates the metric (\ref{thetametric}) becomes 
\bea
g^2 \diff s^2_4 &=& \frac{\diff q^2}{1+ q^2}+q^2\left(\diff\psi^2+\cos^2\psi\diff\varphi_1^2+\sin^2\psi\diff\varphi_2^2\right)~,
\eea
and the gauge field 
\bea
A &=&  \frac{-(1+s\sqrt{1+q^2})\diff\varphi_1 + (s+\sqrt{1+q^2})\diff\varphi_2}{2g\sqrt{(1+s\sqrt{1+q^2})^2+(1-s^2)q^2\cos^2\psi}}~.
\eea
In these coordinates  the ``squashing'' parameter $s$ manifestly parametrizes purely a deformation
of the gauge field  from pure gauge, corresponding to $s=1$. 
A computation shows that the four-dimensional Killing spinor (\ref{epplus}) is constructed from
\bea
\sqrt{y+x} \, \lambda (q,\psi) & =  & \left(\frac{ (s^2-1) + q^2( -\ii \cos\psi + s \sin\psi)^2}{\sqrt{(1-s\sqrt{1+q^2})^2 + q^2(1-s^2)\cos^2\psi }}\right)^{1/2}~,
\label{newspins}
\eea
and in particular this still depends non-trivially on $s$. 

The metric $\diff \tilde s^2_3 $ and gauge field $\tilde A^{(3)}$ 
induced on the conformal boundary defined by $q=\infty$ are 
\bea
\diff \tilde s^2_3 & = & \diff\psi^2+\cos^2\psi\diff\varphi_1^2+\sin^2\psi\diff\varphi_2^2~,\label{newoldmetric}\\
\tilde A^{(3)} & = & -\frac{\tilde h(\psi)}{2g}\left(\diff\varphi_1 -\frac{1}{s}\diff\varphi_2\right)~, 
\label{newoldgauge}
\eea
respectively, where 
\bea
\tilde h^2 (\psi) & = & \frac{s^2}{s^2 \sin^2\psi + \cos^2 \psi}~.
\eea
While the metric is precisely the round metric on the three-sphere, the gauge field is non-trivial and 
takes essentially the same form as in the original coordinates (\ref{boundarygauge}). 

Since the change of coordinates (\ref{newcoc}) is globally smooth, it follows that the boundary metric (\ref{squashedS3}) 
in the original $\theta,\varphi_1, \varphi_2$ coordinates must be in the same conformal class as the round three-sphere metric. One can confirm this by checking that the 
Cotton tensor of the metric (\ref{squashedS3}) vanishes. More explicitly, the change of coordinates
\bea
\cos \psi & = & \frac{s \cos\theta}{h(\theta)} 
\label{bigcoc}
\eea
shows that the two metrics are related by a Weyl rescaling via
\bea
\frac{\diff \theta^2}{h^2(\theta)}+\cos^2\theta\diff\varphi_1^2 + \frac{1}{\s^2}\sin^2\theta\diff\varphi_2^2 \  = \ 
\frac{\tilde h^2(\psi)}{s^2}\left(\diff\psi^2+\cos^2\psi\diff\varphi_1^2+\sin^2\psi\diff\varphi_2^2\right)~,
\eea
while the gauge field correspondingly transforms as
\bea
A^{(3)} & = & -\frac{h(\theta)}{2g} \left(\diff\varphi_1 -\frac{1}{s}\diff\varphi_2\right) 
\ = \ -\frac{\tilde h(\psi)}{2g}\left(\diff\varphi_1 -\frac{1}{s}\diff\varphi_2\right) \ = \ \tilde A^{(3)}~. 
\eea

The Killing spinor on the boundary, which we will denote by $\tilde \chi$, may be extracted by expanding the four-dimensional 
spinor\footnote{In order to do this, one has to note that the change of coordinates (\ref{newcoc}) induces a natural 
change of orthonormal frame adapted to the new radial coordinate $q$.}
determined from (\ref{newspins}) in powers of $q^{1/2}$.  
We find that the three-dimensional spinor is 
\bea
\tilde \chi & = & \begin{pmatrix}  \sqrt{\ii \cos \psi - s \sin \psi }   \\ \sqrt{\ii \cos \psi + s \sin \psi} \end{pmatrix} ~,~~~
\label{funnyspinor}
\eea
and it obeys the following equation\footnote{We used the orthonormal frame defined by $\tilde{e}^1 = \cos\psi \diff\varphi_1,
\tilde{e}^2  = \sin\psi \diff\varphi_2,  \tilde{e}^3  =  -\diff\psi$.} 
\bea
(\tilde \nabla_\alpha - \ii g\tilde A^{(3)}_\alpha) \tilde \chi  +  \frac{\ii \tilde h^2(\psi)}{2s} \tilde \gamma_\alpha 
 \tilde \chi  - \frac{1}{2}\de_\psi \log \tilde h (\psi)  \tilde \gamma_\alpha \sigma_3  \tilde \chi & = & 0~,
\label{funnyeq}
\eea
where $\tilde \nabla_\alpha$ is the connection computed with the round metric in (\ref{newoldmetric}), related to the original one
by $\tilde \nabla_\alpha = \nabla_\alpha - \tfrac{1}{2} \tilde \gamma_\alpha{}^\beta \de_\beta \log \tilde h (\psi)$. 
Note that the spinor (\ref{funnyspinor}) depends on $s$, and it is therefore different from the standard Killing spinors on the round sphere, 
to which it reduces when $s=1$. It may be worth comparing this construction with that  in \cite{Hama:2011ea}: 
here we have a round metric, a gauge field, and a non-standard Killing spinor, whereas in  \cite{Hama:2011ea} they have a squashed metric, a gauge field, 
and a standard Killing spinor. 

Finally, defining a rescaled spinor $\chi$ as 
\bea
\chi &=&  \sqrt{ \frac{\tilde h(\psi)}{s}} \tilde \chi  ~,
\eea
and changing coordinates as in 
(\ref{bigcoc}), the equation (\ref{funnyeq}) 
becomes precisely the Killing spinor equation (\ref{3dKSE}) obeyed by  the original metric (\ref{squashedS3}), gauge field 
(\ref{boundarygauge}), and spinor (\ref{KJKS}). 
We will briefly comment on the field theory 
implication of this in the concluding section. 

\subsection{The holographic free energy}\label{holosec}

In this section we derive an expression for  the holographic free energy of the dual field theory by computing the holographically renormalized
 on-shell action for the gauged supergravity solution (\ref{soln}).  This is a standard application of the prescriptions in the literature
 (see \emph{e.g.} \cite{Skenderis:2002wp}), so the 
reader uninterested in the details may jump to the final formula for the free energy (\ref{holofree}). 

The total renormalized action comprises three types of term: the bulk on-shell action (\ref{4dSUGRA}) is divergent
and therefore one evaluates a regulated action, integrated up to a cut-off $y=r$. Then in general one needs to add  boundary terms appropriate to the 
imposed boundary conditions, and counterterms
that remove the divergent part and give a finite result in the limit $r\to \infty$. 
The general form is therefore
\bea\label{fullaction}
I & = & I_\bk^\gr + I_\bk^\ga  + I_\bd^\gr +  I_\bd^F + I_\ct^\gr +I_\ct^F~,
\eea
where
\bea\label{bulk}
I_\bk^\gr +I_\bk^\ga  &=&  -\frac{1}{16\pi G_4}\int_{B_r} \diff^4x\sqrt{\det g_{\mu\nu}}\left(R[g_{\mu\nu}] + 6g^2 - F^2\right)~,\\\label{GH}
I_\bd^\gr  &=& - \frac{1}{8\pi G_4}\int_{\partial B_r} \dd^3x \sqrt{\det \gamma_{\alpha\beta}} K~,\\\label{ct}
I_\ct^\gr &=&  \frac{1}{8\pi G_4}\int_{\partial B_r} \dd^3x\sqrt{\det \gamma_{\alpha\beta}}  \left(  2g + \frac{1}{2g} R [\gamma_{\alpha\beta}]\right)~,\\
I_\bd^F &= &  I_\ct^F \ = \ 0~.
\eea
Here (\ref{bulk}) is simply the $d=4$, $\mathcal{N}=2$ gauged supergravity action (\ref{4dSUGRA}) with which we started. 
We evaluate this on the solution (\ref{soln}), integrating over the ball $B_r$ that is defined by taking 
$s\leq y\leq r$. The boundary integral (\ref{GH}) is the Gibbons-Hawking term, ensuring that 
the equations of motion (\ref{EOM}) do indeed result from varying the action (\ref{bulk}) 
with fixed boundary metric $\gamma_{\alpha\beta}$ on $\partial B_r\cong S^3$. Here 
$K$ denotes the trace of the second fundamental form of this surface. Finally, 
(\ref{ct}) are the counterterms of reference \cite{Emparan:1999pm}: 
the sum $ I_\bk^\gr   + I_\bd^\gr  $ is divergent as we take the cut-off $r\rightarrow\infty$, 
and the counterterms precisely remove this divergence, giving a finite result for 
(\ref{fullaction}) as $r\rightarrow\infty$.  $R[\gamma_{\alpha\beta}]$ of course denotes 
the Ricci scalar of the induced boundary metric in (\ref{ct}). 

Let us now explain why the boundary term  $I_\bd^\ga$ for the gauge field $A$ is not included in (\ref{fullaction}). 
The AdS/CFT duality requires specifying boundary conditions for fluctuating fields in the bulk.  
In the background of an asymptotically AdS$_4$ metric of the form (\ref{asymptoticmetric}), we 
impose the following boundary condition for the gauge field $A$, in the gauge $A_y=0$, as 
$y\rightarrow \infty$ 
\bea\label{gaugeasymptotic}
A_\alpha &=& A_\alpha^{(3)} + \frac{1}{y} J_\alpha + O\left(\frac{1}{y^2}\right)~.
\eea
This amounts to saying that the gauge field is $O(1)$ to leading order as $y\rightarrow\infty$. 
Notice that our particular gauge field instanton in (\ref{soln}) satisfies (\ref{gaugeasymptotic}). 
Assuming the boundary condition (\ref{gaugeasymptotic}), the variation of the Maxwell action is then easily computed to be
\bea
\delta S_{\mathrm{Maxwell}} &=& -\frac{1}{2}\int_{\mathrm{\bd}} *_3 J\wedge \delta A^{(3)}~.
\eea
Thus  holding $A^{(3)}$ fixed on the boundary leads to a well-defined variational problem  
for the Maxwell equations. In fact this is precisely the boundary condition we shall want, 
since we will be regarding $A^{(3)}$ in (\ref{boundarygauge}) as a fixed background 
gauge field in the next section. With this boundary condition we then do not need to add
a boundary term for the variational problem. Notice from (\ref{gaugeasymptotic}) that the 
Maxwell action is automatically finite, and there is no need for any counterterm for 
$F$.

It is now straightforward  to compute (\ref{bulk}) -- (\ref{ct}), and take the limit $r\rightarrow\infty$ in (\ref{fullaction}). Let us quote the finite contributions. Using the Einstein equation, for the bulk gravity action we obtain 
\bea
I_\bk^\gr & = & \frac{3g^2}{8\pi G_4} \int \dd^4x \sqrt{\det g_{\mu\nu}} \ = \  \frac{3g^2}{8\pi G_4}  \frac{(2\pi)^2}{g^4s (s^2-1)} \int_1^s \dd x \int_s^r \dd y (y^2-x^2) \nonumber\\
& = & \frac{\pi}{2G_4 g^2} + \mathrm{divergent}~,
\label{niufu}
\eea
where the divergent part will be precisely cancelled by the boundary terms.
Curiously, we see that this result is independent of $s$ and indeed it is  exactly the same as that obtained for the round three-sphere. This might have been expected, since the bulk metric is 
just AdS$_4$. However, this expectation is certainly naive, and the result could have depended on $s$ because of the particular slicing of AdS$_4$. 
While it would be interesting to investigate the role of a solution consisting of AdS$_4$ with squashed three-sphere boundary and
 no gauge field instanton, we will not pursue this presently. For the instanton action we compute
\bea
I_\bk^F & = &  \frac{(2\pi)^2}{16\pi G_4} \frac{(s^2-1)}{sg^2} \int_1^s \dd x \int_s^\infty \dd y  \frac{x-y}{(x+y)^3} \ = \ \frac{\pi}{8G_4 g^2}\frac{(s-1)^2}{s}~,
\eea
which is finite as promised and vanishes correctly for $s=1$. 
One can check that  the terms $I_\bd^\gr + I_\ct^\gr$ cancel the divergent part in (\ref{niufu}) and do not contribute a finite part upon taking $r\to \infty$. Combining everything we obtain the finite result
\bea
I &=& \frac{\pi Q^2}{8g^2 G_4}~,
\eea
where we have defined
\bea
Q &\equiv & \frac{\s+1}{\sqrt{\s}} \ = \ b + \frac{1}{b}~, \qquad \mbox{where} \quad  \s \ \equiv \ b^2~.
\eea
We thus obtain the result for the round sphere, for which $s=1$, multiplied by the factor $Q^2/4$.  Note that clearly this result does not depend on the 
choice of coordinates, and thus in particular it applies also to the round sphere boundary metric (plus gauge field).

Finally, setting $g=1$ in order to uplift to eleven-dimensional supergravity via (\ref{uplift}), and using the Newton constant formula in (\ref{Newton}), 
we obtain the gravitational free energy in the Euclidean quantum gravity approximation:
\bea
\label{holofree}
\free_b &=& I \ = \ N^{3/2}Q^2\sqrt{\frac{\pi^6}{8\cdot 27\, \mathrm{Vol}(Y_7)}} \ = \ \frac{Q^2}{4}\, \free_{b=1}~.
\eea
We shall reproduce this formula from a dual large $N$ quantum field theory calculation in the next section.

\section{The field theory side}

\label{fieldtheoryside}

\subsection{Supersymmetric gauge theories on the $U(1)^2$-squashed $S^3$}

In \cite{Hama:2011ea} the authors have constructed ${\cal N}=2$  supersymmetric Lagrangians on a squashed three-sphere with metric 
(\ref{HHLintro}),  for gauge theories comprising Chern-Simons and Yang-Mills terms and matter fields in chiral multiplets. 
They have shown that the Lagrangians and supersymmetry variations 
may be appropriately modified if one includes a background gauge field $A_\alpha$ of the form 
(\ref{boundarygaugeintro}), and the supersymmetry parameter\footnote{This was denoted $\epsilon$ in \cite{Hama:2011ea}; we hope this will not generate confusion.} 
$\chi$ obeys the modified Killing spinor equation
\bea\label{KSEagain}
(\nabla_\alpha - \ii  A_\alpha)\chi - \frac{\ii}{2f(\theta)}\gamma_\alpha\chi &=& 0~.
\eea
Although the construction of \cite{Hama:2011ea} appears to require the existence of a ``second'' Killing spinor, denoted $\bar\epsilon$ there, in fact this is simply
the charge conjugate $\chi^c$, which in general satisfies the same Killing spinor 
equation (\ref{KSEagain}) but with $A_\alpha$ replaced by $-A_\alpha$.
In the following we will summarize the supersymmetric Lagrangians constructed by  HHL, and their computation of the partition function using localization. 
For simplicity we will consider a single vector multiplet $V$ and a single chiral multiplet $\Phi$, transforming in the fundamental representation of the gauge group. 

A 3d ${\cal N} = 2$ vector multiplet $V$ consists of a gauge field $\mathscr{A}_\alpha$, a scalar field $\sigma $, a two-component Dirac spinor $\lambda$, 
and scalar field $D$, all transforming in the adjoint representation of the gauge group. 
The matter field $\Phi$ is a chiral multiplet, consisting of a complex scalar $\phi$, a fermion $\psi$ and an auxiliary scalar $F$, which 
we take here to be in the fundamental representation of the gauge group. This is assumed to have an arbitrary R-charge
$\Delta$. The ${\cal N} = 2$ Lagrangian constructed in  \cite{Hama:2011ea}  consists of three
terms $S = S_{\mathrm{CS}} + S_\mathrm{mat}+ S_{\mathrm{YM}}$, that we  discuss in 
turn.\footnote{Instead of the  conventions of \cite{Hama:2011ea}, we will adopt a somewhat more standard notation.}
The Chern-Simons term is unchanged with respect to the  
expression in flat space and reads
\bea\label{SCS}
S_{\mathrm{CS}} & =  & \frac{k}{4\pi} \int \mathrm{Tr} 
\left[ \mathscr{A} \wedge \dd \mathscr{A}  -\frac{2\ii}{3} \mathscr{A}  \wedge \mathscr{A} \wedge \mathscr{A}   
-  * \mathbf{1} (\lambda^\dagger \lambda  - 2 D \sigma  ) \right]~,
\eea
where $k$ is the integer Chern-Simons level. 
The matter Lagrangian reads 
\begin{eqnarray}
S_\text{mat}
&=& \int \dd^3 x \sqrt{\det \gamma_{ij}} \bigg[
\cov_\alpha \phi^\dagger \cov^\alpha \phi
+\phi^\dagger \sigma^2\phi
+\ii\phi^\dagger  D \phi  
+F^\dagger F  \nonumber\\[2mm]
&&~~~~~~~~~~ -\ii\psi^\dagger \gamma^\alpha \cov_\alpha \psi
+\ii\psi^\dagger \sigma\psi
+\ii\psi^\dagger \lambda\phi
 -\ii\phi^\dagger \lambda^\dagger \psi
 \nonumber \\[2mm] 
&&~~~~~~~~~~  +\ii\phi^\dagger \frac{\sigma}f\phi
+\frac{2\ii(\Delta-1)}f v^\alpha \cov_\alpha\phi^\dagger \phi 
 +\frac{\Delta(2\Delta-3)}{2f^2}\phi^\dagger \phi
 +\frac \Delta 4 R \phi^\dagger \phi
\nonumber\\[2mm]
&&~~~~~~~~~~ -\frac{1}{2f}\psi^\dagger \psi
 +\frac{\Delta-1}f\psi^\dagger \gamma^\alpha v_\alpha \psi
\bigg]~.
\end{eqnarray}
The first two lines reduce  to 
the usual expressions in flat space (and $A_\alpha=0$), 
while the last two lines are new terms necessary for supersymmetry in the curved background.  
Here $R$ denotes the scalar curvature of the background metric, and $v^\alpha$ is the vector bilinear 
$v^\alpha \equiv \chi^\dagger  \gamma^\alpha \chi $
constructed from the spinor $\chi$, normalized so that $\chi^\dagger\chi =1$, and satisfying $v^\alpha v_\alpha=1$.
The covariant derivative is defined as 
\bea
\cov_\alpha & = &  \nabla_\alpha - \ii [\mathscr{A}_\alpha, \cdot ] - \ii  \Delta A_\alpha~,
\eea
where $\nabla_\alpha$ is the metric covariant derivative, $\mathscr{A}_\alpha$ is the gauge field and  $A_\alpha$ is the background $U(1)$ gauge field. 
$\Delta$ is the R-charge (or conformal dimension in the conformally invariant case) of the field on which $\cov_\alpha$ acts. 
This Lagrangian is invariant under a set of supersymmetry variations, independently of the function $f$ \cite{Hama:2011ea}; however we will not write these
here.  Notice that although in Euclidean signature one can have two independent 
supersymmetry parameters, denoted $\epsilon$ and $\eta$ in  \cite{Kapustin:2009kz}, in the construction of \cite{Hama:2011ea} they are related: the second spinor is simply 
the charge conjugate of the first, as we have already noted. 
Finally, the Yang-Mills Lagrangian reads
 \begin{eqnarray}
S_\text{YM}  
 &=&  \frac{1}{g_{\mathrm{YM}}^2} \int \dd^3 x \sqrt{\det \gamma_{ij}} ~ \mathrm{Tr}\bigg[
  \frac14 \mathscr{F}_{\alpha\beta}  \mathscr{F}^{\alpha\beta}+\frac12\cov_\alpha\sigma \cov^\alpha\sigma+\frac12\left(D+\frac\sigma f\right)^2 
\nonumber \\ 
&& ~~~~~~~~~~~
 +\frac \ii 2\lambda^\dagger \gamma^\alpha \cov_\alpha\lambda
+\frac \ii 2\lambda^\dagger [\sigma,\lambda]
 - \frac1{4f}\lambda^\dagger\lambda
 \bigg]~,
\end{eqnarray}
where notice that the bosonic part is positive semi-definite, and hence the Yang-Mills Lagrangian  acts as a regulator in the path integral.
This will be important for the localization argument. 
For an Abelian gauge group there exists also a supersymmetric version 
of the FI parameter; however this is not relevant for the application in this paper. 

\subsection{Localization of the partition  function}
\label{localsection}

The supersymmetric Yang-Mills and matter Lagrangians above are in fact total supersymmetry variations with respect to the supersymmetry $\delta_{\chi^c}$ 
generated by $\chi^c$ (of course one could just swap the definitions of $\chi^c$ and $\chi$), and therefore they can be used for applying localization.
In particular, we have that
\bea
{\cal L}_\text{YM}  & = & \delta_{\chi^c} \left( \delta_{\chi} \, \mathrm{Tr}\left(\frac{1}{2}\lambda^\dagger \lambda - 2 D \sigma\right)\right)~,\nn\\
{\cal L}_\text{mat}  & = & \delta_{\chi^c} \left( \delta_{\chi} \, \mathrm{Tr}\left(\frac{1}{2}\psi^\dagger \psi - 2 \ii \phi^\dagger \sigma \phi\right)\right)~.
\eea
Therefore both these terms may be included  in the partition function multiplied by arbitrary parameters, so that  
the total (Euclidean) partition function of a Chern-Simons(-Yang-Mills)-matter theory may be written  as
\bea
Z & = & \int {\cal D}[\mathrm{all~fields}]\, \me^{ - S_{\mathrm{CS}}   - t S_{\mathrm{YM}} - (t+1) S_\mathrm{mat}}~,
\eea
and by the standard localization argument this is independent of the parameter $t$. The physical theories we are interested in correspond to the value $t=0$, 
whereas in the limit $t\to \infty$ all the contribution comes from the saddle-point, which is a supersymmetric configuration of fields in the curved background.
This is characterized by all fields vanishing, except the scalar fields in the vector multiplet which satisfy 
\bea
  f D  &  = & - \sigma \  = \  \mathrm{constant}~.  
\label{locca}
\eea
Notice that $\sigma$ is a matrix-valued constant field, while $D$ is not constant and depends on $f$. However, we will see that this dependence will disappear 
completely from the final answer.

The partition function  receives a classical contribution from the Chern-Simons action $S_{\mathrm{CS}}$ (\ref{SCS}) evaluated on the solution  
(\ref{locca}), and a one-loop contribution from the Gaussian integral over quadratic fluctuations of all the fields (bosonic and fermionic)
in $ S_\mathrm{mat} + S_\mathrm{YM}$, around the classical solution (\ref{locca}). 
The key observation of the authors of \cite{Hama:2011ea} is that the bosonic and fermionic eigenmodes entering the one-loop determinants 
are paired by supersymmetry, and therefore their detailed form is irrelevant since they give cancelling contributions.
One can thus circumvent a detailed computation of the spectrum of the relevant kinetic operators 
by identifying the few  eigenmodes that do not pair, and therefore give a net contribution to the one-loop determinant.

Before describing the details, and our main aim of deriving (\ref{Fintro}), let us note that our key observation 
here is that essentially all  the computations in section 5 of \cite{Hama:2011ea} go through independently 
of the specific functional form of $f(\theta)$ entering the metric (\ref{HHLintro}). In fact one needs only that 
$f(\theta)$ enters the Killing spinor equation as in (\ref{3dKSEintro}), and the gauge field as in (\ref{boundarygaugeintro}), 
together with the boundary conditions $|f(\theta)|\to 1/b^2$ as $\theta \to 0$ and $|f(\theta)|\to 1$ as $\theta \to \tfrac{\pi}{2}$, 
which ensure regularity of the metric. Recall that for the particular ellipsoid metric in \cite{Hama:2011ea} 
one has $f(\theta)=\sqrt{\sin^2\theta+\tfrac{1}{b^4}\cos^2\theta}$, while 
our ``hyperbolic ellipsoid'' satisfies the same equations but with $f(\theta)=-1/h(\theta) = -1/\sqrt{b^4\cos^2\theta+\sin^2\theta}$.
Having emphasized this, we now briefly summarize the steps in section 5 of \cite{Hama:2011ea}, and how these results then lead to the 
partition function given by (\ref{partfun}), (\ref{Fintro}).

We consider first a chiral matter multiplet $\Phi=(\phi,\psi,F)$, which for simplicity we assume has unit charge under 
a single Abelian vector multiplet -- the extension to arbitrary representations of a non-Abelian 
gauge group is straightforward, and we will write the result relevant for quiver theories at the 
end of the section. In this set-up, it is simple to verify that there is a pairing between eigenmodes 
of the scalar kinetic operator for $\phi$ and the spinor kinetic operator for $\psi$. More precisely, a scalar eigenmode with eigenvalue 
$\mu(\mu-2\ii\sigma)$ is paired with two spinor eigenmodes with eigenvalues $\mu$, $2\ii\sigma-\mu$. Here 
$\sigma$ is the scalar in the vector multiplet under which $\Phi$ has unit charge, which is constant and satisfies (\ref{locca}). This pairing  involves contractions or products with
the Killing spinor $\chi$, and the above statements then depend only on the Killing spinor equation (\ref{3dKSEintro}) and the identity 
$v_\alpha \gamma^\alpha \chi = \chi$, but not on the specific expression for $f(\theta)$.  
The contributions of the paired modes to the partition function then precisely cancel, as is familiar in supersymmetric theories.

Thus we need only consider the modes that do \emph{not} have a superpartner under the above pairing. The first such class of modes are
spinor eigenmodes characterized by having zero inner product with the Killing spinor $\chi$, so that the corresponding scalar in the would-be pairing is identically zero. 
One finds that the eigenvalues of such modes are
\bea\label{spinormode1}
\mu &=& \ii\sigma + m + nb^2 - \tfrac{1}{2}(\Delta-2)(1+b^2)~,
\eea
where the eigenfunction has charge $(m,-n)$ under $\partial_{\varphi_1}$, $\partial_{\varphi_2}$, so $m,n\in\Z$. 
The dependence of the modes on the coordinate $\theta$ in turn depends on the function $f(\theta)$. However, 
the \emph{normalizability} depends only on the \emph{boundary conditions} of $f(\theta)$ at $\theta=0$, $\theta=\pi/2$, and
this is determined by regularity of the metric. The upshot is that the modes (\ref{spinormode1}) 
are normalizable if and only if $m,n\geq 0$, precisely as in \cite{Hama:2011ea}.
The second class of modes are where the two spinor eigenmodes associated to a given scalar 
are linearly dependent. In this case one finds the spectrum
\bea\label{spinormode2}
\mu &=& \ii\sigma - m - nb^2 - \tfrac{1}{2}\Delta(1+b^2)~,
\eea
where again normalizability requires $m,n\geq 0$. The first type of spinor modes (\ref{spinormode1})  are left uncancelled by the 
scalar determinant, while the second type of spinor modes (\ref{spinormode2}), while paired with a scalar, will then be double counted. 
Thus the first contribute to the numerator, while the second effectively contribute to the denominator in the 
 one-loop determinant of the chiral multiplet, giving
\bea
 Z_{\mathrm{one-loop}}^{\mathrm{mat}}(\sigma) &=& 
 \frac{\text{det}\, \Delta_\psi}
      {\text{det}\, \Delta_\phi}
  \ = \  \prod_{m,n\ge0}
 \frac{mb^{-1}+nb+\frac Q2+\ii\tfrac{1}{b}\sigma+\frac Q2(1-\Delta)}
      {mb^{-1}+nb+\frac Q2-\ii\tfrac{1}{b}\sigma-\frac Q2(1-\Delta)}\nn\\
 & =& s_b\big(\tfrac{\ii Q}2(1-\Delta)-\tfrac{1}{b}\sigma\big)~,
\label{nicedet1}
\eea
where recall that $Q=b+1/b$, and $s_b$ is by definition the double sine function.

The analysis of the one-loop determinant of the vector multiplet $V=(\mathscr{A}_\alpha,\sigma,\lambda,D)$, for an arbitrary gauge group $G$, is very similar. In this case, after gauge fixing and combining with the volume of the gauge group, only 
the transverse vector eigenmodes contribute to the one-loop determinant. In this case the transverse vector eigenmodes are paired  
with superpartner spinor eigenmodes, both of the same eigenvalue $\mu$. The unpaired modes, which then contribute
to the partition function, again fall into two classes. The first are spinor eigenmodes for the kinetic operator for $\lambda$ that pair with 
identically zero vector eigenmodes. These have eigenvalues 
\bea\label{vectormode}
\mu &=&  m+nb^2+\ii\alpha(\sigma)~,
\eea
where $\alpha$ runs over the roots of $G$.
Again normalizability requires $m,n\geq 0$, but not \emph{both} zero, {\it i.e.} the mode 
 $m=n=0$ is \emph{not} a normalizable unpaired spinor mode.
The second are vector eigenmodes that pair with identically zero 
spinor eigenmodes. These also have eigenvalues (\ref{vectormode}), but now normalizability requires 
$m,n\leq -1$. The first class then contribute to the numerator, while the second contribute to the denominator in the 
 one-loop determinant of the vector multiplet, giving a total contribution
\bea\label{detvectornice}
 Z_{\mathrm{one-loop}}^{\mathrm{vector}}(\sigma)&= &  \frac{\text{det}\, \Delta_\lambda}
      {\ \text{det}\, \Delta_{\mathscr{A}_\alpha^\perp}} \nn\\
      & = &  \prod_{\mathrm{roots}\, \alpha}\frac{1}{\ii \alpha(\sigma)}\prod_{m,n\geq 0}\frac{m+nb^2+\ii\alpha(\sigma)}{-m-1+(-n-1)b^2+\ii\alpha(\sigma)}\nn\\
      &=& \prod_{\mathrm{positive\, roots}\, \alpha}\frac{4\sinh(\pi\alpha( \sigma))\sinh(\pi b^{-2}\alpha(\sigma))}{\alpha(\sigma)^2} ~.
\eea
Notice here we have included the $m=n=0$ mode in (\ref{vectormode}) in the numerator, but then explicitly 
divided by $\ii \alpha(\sigma)$ to remove it in the middle line of (\ref{detvectornice}). The equality in the 
last line is explained in appendix \ref{vectorapp}.

In fact we shall be interested only in the case where $G=U(N)$. In this case we may take the Cartan to be the diagonal $N\times N$ matrices, 
and write
\bea
\sigma &=& \left(\frac{\lambda_1}{2\pi},\ldots,\frac{\lambda_N}{2\pi}\right)~,
\eea
where $\frac{\lambda_i}{2\pi}$, $i=1,\ldots, N$, are the eigenvalues of $\sigma$. Then the roots of $G$ are labelled by integers $i\neq j$ with
\bea
\alpha_{ij}(\sigma) &=& \frac{\lambda_i-\lambda_j}{2\pi}~,
\eea
with a choice of positive roots being $\{\alpha_{ij}\mid i<j\}$.
Taking into account also the Vandermonde determinant (see appendix \ref{vectorapp}), the one-loop vector multiplet determinant (\ref{detvectornice}) then reduces to
\bea
 \prod_{i<j} 4\sinh \frac{\lambda_i-\lambda_j}{2}\sinh \frac{\lambda_i-\lambda_j}{2b^2}~,
\eea
which is of the form presented in (\ref{Fintro}). 

For a chiral multiplet $\Phi$ in a general representation $\mathcal{R}$ of the gauge group $G$, one should simply replace 
$\sigma$ in (\ref{nicedet1}) by $\rho(\sigma)$, and then take the product over weights $\rho$ in a weight-space decomposition 
of $\mathcal{R}$. For the 
 bifundamental representation of $U(N)_I\times U(N)_J$, this is
 \bea
 \rho_{ij}(\sigma) &=&  \frac{\lambda^I_i-\lambda^J_j}{2\pi}~,
 \eea
 which again directly leads to the form presented in (\ref{Fintro}).
 
 Finally, the first term in (\ref{Fintro}) is the contribution from the classical Chern-Simons action, which 
 upon localization reads
 \bea\label{Scl}
S_{\mathrm{CS}} &=& \frac{\ii k}{4\pi}\int_{S^3_{\mathrm{squashed}}} 2\, \mathrm{Tr} (D\sigma)\nn\\
 & = & -\frac{\ii k}{4\pi}  \int_{\theta=0}^{\pi/2}\int_{\varphi_1=0}^{2\pi}\int_{\varphi_2=0}^{2\pi} \sqrt{\det \gamma_{ij}}\, \diff^3x \frac{2}{f(\theta)} \mathrm{Tr} \, \sigma^2 \nn\\
 & =& - \frac{\ii k}{4\pi b^2}\sum_{i=1}^N\lambda_i^2~.
 \eea
 Here we have substituted $D=-\sigma/f$ (\ref{locca}), used the Riemannian measure $\sqrt{\det \gamma_{ij}} = \tfrac{1}{b^2}f(\theta)\sin\theta\cos\theta$ 
 for the metric (\ref{HHLintro}), so that 
 $f(\theta)$ cancels in (\ref{Scl}), and substituted $\mathrm{Tr}\, \sigma^2 = \sum_{i=1}^N\left( \frac{\lambda_i}{2\pi}\right)^2$. 
 This completes our derivation of the partition function (\ref{partfun}), (\ref{Fintro}).

\subsection{Large $N$ limit of the free energy}

In this section we evaluate the partition function (\ref{partfun}), for a large class of Chern-Simons quiver theories, in the ``M-theory limit'' 
in which the rank $N$ is taken to infinity while the Chern-Simons levels $k_I$ are held fixed. This is a relatively straightforward modification of the 
computation presented in \cite{Martelli:2011qj, Cheon:2011vi, Jafferis:2011zi}, and so we shall be as brief as possible.\footnote{Very recently 
we note that a completely different method has been found for computing this M-theory limit \cite{Marino:2011eh}.}

As in \cite{Herzog:2010hf}, the idea is to compute the integral (\ref{partfun}) in a saddle point approximation. 
Solutions to the saddle point equations  may be viewed as zero force configurations between the eigenvalues $\lambda_i^I$, which interact via 
a potential. As the number 
of eigenvalues $N$ for each gauge group tends to infinity, one has a continuum limit in which one can replace the sums over eigenvalues in (\ref{partfun}) 
by integrals. In particular, one can then separate the interactions between eigenvalues into ``long range forces'', 
for which the interaction between eigenvalues is non-local, plus a local interaction. A key point, observed in \cite{Martelli:2011qj}, 
is that for an appropriate class of \emph{non-chiral} Chern-Simons quiver theories, these long ranges forces automatically cancel. We begin 
by showing that this statement is unmodified for the corresponding supersymmetric theories on the squashed 
sphere, with $b\neq 1$.

The long range forces referred to above are related to the leading terms in an asymptotic expansion of the functions 
appearing in the integrand in (\ref{partfun}). In particular, if we define 
\bea
f_b(z) &\equiv & \log s_b(z)~,
\eea
where $s_b(z)$ is the double sine function, then the long range forces are determined by 
\bea\label{fass}
f_b^{\mathrm{asymp}}(z) & \equiv & \frac{\ii \pi}{2}\left(z^2+\frac{b^2+b^{-2}}{12}\right)\mathrm{sign} \left(\real z\right)~.
\eea
Here $f_b(z)-f_b^{\mathrm{asymp}}(z)$ has the property that it tends to zero as $|\real z|\rightarrow\infty$ 
\cite{Kharchev:2001rs, Bytsko:2006ut}. Similarly, we have
\bea\label{sass}
\left[\log \sinh {z}\right]^{\mathrm{asymp}} &\equiv & z\, \mathrm{sign} \left(\real z\right)~.
\eea

One then takes the continuum limit of (\ref{partfun}), so that the sums become Riemann integrals
\bea
\frac{1}{N}\sum_{i=1}^N &  \longrightarrow & \int_{\xmin}^{\xmax} \rho(x)  \diff x~,
\eea
where we make the following ansatz for the eigenvalues \cite{Herzog:2010hf}
\bea\label{eigen}
\lambda^I(x) &=& N^\alpha x + \ii y^I(x)~,
\eea 
with $\alpha>0$.
Note here that we have deformed the real eigenvalues
in (\ref{partfun}) into the complex plane in (\ref{eigen}), as is often necessary when performing the 
saddle point method for evaluation of integrals, and that  
the function $\rho(x)$ describes the eigenvalue density.  In this limit, and substituting the functions 
$f_b(z)$ and $\sinh z$ by their asymptotic forms in (\ref{fass}), (\ref{sass}), we obtain 
the following long range contribution to $F$:
\bea\label{Fasym}
-F_{\mathrm{asymp}}&  =  & N^2\int_{\xmin}^{\xmax} \rho(x)\diff x\int_{\xmin}^{\xmax} \rho(x')\diff x' \, \mathrm{sign}(x-x')
\Bigg\{\frac{Q}{4b}\sum_{I=1}^G \lambda^I(x)-\lambda^I(x') \nn\\
&& - \sum_{I\rightarrow J}\frac{Q}{4b} (1-\Delta_{I,J})\left[\lambda^I(x)-\lambda^J(x')\right]
+ \frac{\ii \pi}{2b^2}\left(\frac{\lambda^I(x)-\lambda^J(x')}{2\pi}\right)^2\Bigg\}~.
\eea
Here we have already used the fact that a constant inserted into the curly bracketed expression 
in (\ref{Fasym}) does not affect the integral, due to the skew symmetry under exchanging 
$x\leftrightarrow x'$. In fact this same symmetry may then be used to argue that the last quadratic 
term in (\ref{Fasym}) also contributes zero, provided that the quiver is \emph{non-chiral}: that is, 
for every bifundamental field transforming as $I\rightarrow J$, there is an associated field 
transforming as $J\rightarrow I$. The terms quadratic in $\lambda^I$ in (\ref{Fasym}) then cancel pairwise, 
and we may further simplify (\ref{Fasym}) to
\bea
-F_{\mathrm{asymp}} & = & \frac{QN^2}{2b}\int_{\xmin}^{\xmax} \rho(x)\diff x\int_{\xmin}^{\xmax} \rho(x')\diff x' \, \mathrm{sign}(x-x')
\Bigg\{\sum_{I=1}^G \lambda^I(x) \nn\\ && - \frac{1}{2} \sum_{I\rightarrow J} (1-\Delta_{I,J})\left[\lambda^I(x)+\lambda^J(x)\right]\Bigg\}~.
\eea
The coefficient of $\lambda^I(x)$ in the integrand is then
\bea\label{beta}
1 - \frac{1}{2}\sum_{\mathrm{fixed}\, I\rightarrow J}(1-\Delta_{I,J}) - \frac{1}{2}\sum_{\mathrm{fixed}\, I\leftarrow J}(1-\Delta_{J,I})~.
\eea
Thus provided this expression vanishes for each $I$, the long range contribution $F_{\mathrm{asymp}}$ is zero. As noted in 
\cite{Martelli:2011qj}, curiously (\ref{beta}) are in fact the beta function equations for the parent four-dimensional 
$\mathcal{N}=1$ quiver gauge theory. 

We thus now restrict to non-chiral Chern-Simons quiver gauge theories with an R-symmetry 
that satisfies (\ref{beta}). For such theories the long range forces between eigenvalues 
cancel, and it remains to compute the leading order contribution to the free energy 
in the M-theory limit. From (\ref{partfun}) one easily computes
\bea\label{classicalF}
F_{\mathrm{classical}} &=& \frac{N^{1+\alpha}}{2\pi b^2}\int_{\xmin}^{\xmax}\rho(x)\diff x \sum_{I=1}^G k_I x y^{I}(x) + o(N^{1+\alpha})~,
\eea
so that the $b=1$ result is simply rescaled by $1/b^2$. The one-loop contribution from each vector multiplet is
\bea\label{gaugeF}
F_{\mathrm{gauge}} &=& \frac{\pi^2 b Q N^{2-\alpha}}{6}\int_{\xmin}^{\xmax} \rho(x)^2\diff x + o(N^{2-\alpha})~,
\eea
leading instead to a $bQ/2$ rescaling of the $b=1$ result. Notice that in obtaining (\ref{gaugeF}) we are effectively using the 
substitution 
\bea
\log \sinh z -\left[\log \sinh {z}\right]^{\mathrm{asymp}}  &\simeq &- \frac{\pi^2}{6}\delta (\real z)~,
\eea
in (\ref{partfun}) -- a more detailed discussion of precisely how this delta function arises may be found around equation (3.33) of 
\cite{Martelli:2011qj}. Finally, the one-loop matter contribution follows from the similar 
approximation (see also appendix A of \cite{Imamura:2011wg})
\bea
f_b(z)-f_b(z)^{\mathrm{asymp}} & \simeq & \frac{\pi}{3}\delta(\real z) \left[(\imag z)^3-\tfrac{1}{4}(b^2+b^{-2})\imag z\right]~,
\eea
which for a single bifundamental field $I\rightarrow J$ then gives
\bea
F_{ I, J} \, =\,  -\frac{2\pi^2 b N^{2-\alpha}}{3}\int_{\xmin}^{\xmax} \rho(x)^2\diff x \Big[Y_{I, J}(x)^3- 
 \tfrac{1}{4}(b^2+b^{-2})Y_{I, J}(x)\Big]+o(N^{2-\alpha})~,
\eea
where we have defined
\bea
Y_{I, J}(x) &\equiv & \frac{Q}{2}(1-\Delta_{I,J}) - \frac{y^I(x)-y^J(x)}{2\pi b}~.
\eea

Now, the sum over $G$ $U(N)$ vector multiplets gives $G$ times the contribution (\ref{gaugeF}). Using (\ref{beta}) we may then write
\bea
G &=& \sum_{I\rightarrow J} (1-\Delta_{I,J})~,
\eea
where the sum is over all bifundamental fields. Using the fact that the quiver is non-chiral, with each bifundamental $I\rightarrow J$ being 
paired with a corresponding bifundamental $J\rightarrow I$, the contributions from the 
one-loop vector and matter multiplets combine to give
\bea\label{oneloopF}
F_{\mathrm{one-loop}} &=& \frac{ (bQ)^3 \pi^2 N^{2-\alpha}}{2^3\cdot 3b^2} \int_{\xmin}^{\xmax} \rho(x)^2\diff x\sum_{\mathrm{pairs}\, I\leftrightarrow J} 
\frac{\left(2-\Delta^+_{I,J}\right)}{2}\Bigg\{\Delta^+_{I,J}(4-\Delta^+_{I,J}) \nn\\
&& - 3 \left[\frac{2\left(y^I(x)-y^J(x)\right)}{\pi bQ}+\Delta^-_{I,J}\right]^2\Bigg\} +o(N^{2-\alpha})~,
\eea
where we have defined
\bea
\Delta^\pm_{I,J} & \equiv  & \Delta_{I,J}\pm \Delta_{J,I}~,
\eea
for each bifundamental pair.

As in the $b=1$ case, we thus see that in order for the classical and one-loop contributions in (\ref{classicalF}), (\ref{oneloopF}) 
to be the same order in $N$, which in turn is necessary for a saddle point solution, we must take $\alpha=\tfrac{1}{2}$. 
Then making the change of variable
\bea
\hat{y}^I(x) &\equiv & \frac{2}{bQ}y^I(x)~,
\eea
the leading order action obtained by combining the classical and one-loop terms is
\bea\label{totalF}
F &=& N^{3/2}\Bigg\{\frac{bQ}{2b^2}\int_{\xmin}^{\xmax}\rho(x)\diff x \left[\sum_{I=1}^G \frac{k_I}{2\pi} x \hat{y}^{I}(x)\right] + \frac{(bQ)^3}{2^3 b^2} \frac{\pi^2}{3} \int_{\xmin}^{\xmax} \rho(x)^2\diff x  \\ 
&&\sum_{\mathrm{pairs}\, I\leftrightarrow J}  \frac{\left(2-\Delta^+_{I,J}\right)}{2}\Bigg\{\Delta^+_{I,J}(4-\Delta^+_{I,J})  - 3 \left[\frac{\hat{y}^I(x)-\hat{y}^J(x)}{\pi }+\Delta^-_{I,J}\right]^2\Bigg\}\Bigg\}\nn
~.
\eea
Setting $b=1$ we precisely recover the results of \cite{Martelli:2011qj, Cheon:2011vi, Jafferis:2011zi}. For $b>1$ we see 
that the classical contribution has effectively been scaled by $bQ/2b^2$, while the one-loop contribution has been scaled 
by $(bQ)^3/2^3b^2$, relative to the $b=1$ result. Alternatively, we may view this as rescaling the \emph{entire} 
action by the latter factor of $(bQ)^3/2^3b^2$, and in turn rescaling the \emph{Chern-Simons couplings} $k_I$ 
by $k_I\rightarrow  (2/bQ)^2 k_I$. Provided the Chern-Simons quiver theory is dual to 
M-theory on an AdS$_4\times Y_7$ background, then the free energy in the $b=1$ case 
scales as $\sqrt{k}$ if one multiplies $k_I\rightarrow k\cdot k_I$, since the volume of $Y_7$ scales as $1/k$. 
Taking this into account, we see from (\ref{totalF}) that the final result for the free energy, obtained by 
extremizing (\ref{totalF}) and evaluating at the critical point, is given by
\bea
\mathcal{F}_b &=& F_{\mathrm{critical}} \ = \ \frac{(bQ)^3}{2^3b^2}\cdot \frac{2}{bQ}\cdot \mathcal{F}_{b=1} \nn \\ 
&=& \frac{Q^2}{4}  \mathcal{F}_{b=1}~.
\eea
The large $N$ matching of the free energy on the round three-sphere, $\mathcal{F}_{b=1}$, with the holographic free energy 
computed in AdS$_4$ was first demonstrated in \cite{Drukker:2010nc} for the ABJM model, and extended to larger classes of theories in  
\cite{Herzog:2010hf,Santamaria:2010dm,Martelli:2011qj,Cheon:2011vi,Jafferis:2011zi}.
Thus we precisely reproduce the dual gravity computation (\ref{holofree}).

\section{Discussion}

\label{discussionsex}

In this paper we presented a class of supersymmetric solutions of eleven-dimensional supergravity, and conjectured that this is dual to supersymmetric 
${\cal N}=2$ gauge theories on the background of a squashed three-sphere and a $U(1)$ gauge field, whose partition function 
may be computed using supersymmetric localization \cite{Hama:2011ea}. Indeed, although the restriction of our gravity solution 
to the three-dimensional boundary is slightly different to the background considered in \cite{Hama:2011ea}, we have
nevertheless argued that the localized partition functions for the two backgrounds are equal.
Recall that in section \ref{roundsec} we showed that our particular squashed $S^3$ is related by a 
smooth Weyl transformation to the round $S^3$. This is particularly clear from the gravity dual description, 
where the two metrics simply arise from different slicings of AdS$_4$. However, what's not so clear 
is whether the localization and field theory partition function are invariant under Weyl rescalings, although 
we expect that this will be true. At least 
for large $N$, this would necessarily have to be true from the AdS/CFT correspondence. The possibility of obtaining the gravity dual 
of exactly the field theory background in  \cite{Hama:2011ea}, or for other choices of the function $f(\theta)$, remains an open problem. 

As a  non-trivial test of this correspondence we have successfully 
matched the holographic free energy to the large $N$ behaviour of the field theoretic free energy, computed from the matrix model.
On both sides the result takes the form of that of the round three-sphere result, multiplied by the factor $(Q/2)^2$ where $Q=b+1/b$.
One of the original motivations for studying supersymmetric gauge theories on the squashed $S^3$ in \cite{Hama:2011ea} 
was the relation, via the AGT correspondence \cite{Alday:2009aq}, to Liouville or Toda theories with coupling $b$. 
Of course, a major difference here is that our $N$ counts the number of M2-branes, while in the 
AGT correspondence it is M5-branes that appear.

This construction potentially has numerous generalizations. On the one hand one should explore the possibilities for curved backgrounds 
on which one can place rigid supersymmetric field theories, pursuing the work of \cite{Festuccia:2011ws}. On the other hand, it is then natural to 
attempt to construct gravity duals for each of these cases. Indeed, the relation between rigid and local supersymmetry is already
clear from the results of \cite{Festuccia:2011ws}. We anticipate that immediate generalizations will arise from the class of Plebanski-Demianski solutions 
to four-dimensional ${\cal N}=2$ gauged supergravity \cite{AlonsoAlberca:2000cs}, or indeed from yet more general (Euclidean) supersymmetric
solutions to this theory  \cite{Caldarelli:2003pb,Cacciatori:2004rt,Dunajski:2010zp,Dunajski:2010uv}. For example, the gravity dual to the construction 
in \cite{Imamura:2011wg} might be found within these classes. Another immediate extension is to embed (via a consistent truncation) 
our solution, and these generalizations, in the context 
of general ${\cal N}=2$,  AdS$_4\times Y_7$ solutions that the authors have investigated in  \cite{Gabella:2011sg}.
It would also be natural to explore gauge/gravity dualities where the field theory lives on 
non-trivial curved backgrounds in dimensions other than three. In particular, 
we expect that this point of view should be useful for constructing supersymmetric gauge theories
on deformed four-spheres or other curved four-manifolds. 

\subsection*{Acknowledgments}
\noindent
We thank Fernando Alday and Nadav Drukker for useful discussions and comments. J.~F.~S.  and D.~M.  would like to thank the 
Centro de ciencias de Benasque Pedro Pascual for hospitality,
and for the stimulating discussions with participants that gave rise to  this work.
J.~F.~S. would like to thank the Simons Center for Geometry and Physics for hospitality 
while part of this research was carried out. D.~M. is supported by an EPSRC Advanced
 Fellowship EP/D07150X/3, A.~P. by an STFC grant and an A.G. Leventis Foundation grant, and
J.~F.~S. by a Royal Society University Research Fellowship. 

\appendix

\section{Plebanski-Demianski origin of the solution}
\label{PDappendix}

The Plebanski-Demianski solutions \cite{PD} are a large class of exact solutions to four-dimensional 
Einstein-Maxwell theory, {\it i.e.} they solve the equations of motion (\ref{EOM}). In fact they are 
the \emph{most general} such solutions of Petrov type D, and it is this property that allows 
one to solve the Einstein equations in closed form. Many well-known solutions, such 
as the Kerr-Newman solution describing a rotating, charged black hole, arise as particular limits.

Our starting point will be the form of the Plebanski-Demianski solutions essentially as  presented in \cite{AlonsoAlberca:2000cs}. 
In Euclidean signature, the metric can be written
\bea\label{PDmetric}
\dd s^2_4  &= & \frac{\q}{q^2- p^2} (\dd\tau + p^2 \dd\sigma)^2 + \frac{\p}{p^2-q^2} (\dd\tau + q^2\dd\sigma )^2   +\frac{q^2-p^2 }{\q } \dd q^2 \nonumber\\ 
&&+   \frac{p^2- q^2 }{\p }\dd p^2 ~,
\eea
where $\p$ and $\q$ are quartic polynomials given by\footnote{Note that the constant $Q$ defined in this appendix is different from the parameter 
$Q=b+1/b$ discussed in the main text. To obtain the metrics in the Euclideanized form presented here, one should take 
the solutions as presented in \cite{AlonsoAlberca:2000cs} and map
$q\mapsto \ii q$, $M\mapsto \ii M$, $Q\mapsto \ii Q$ (together with $\sigma\mapsto-\sigma$).}
\bea\label{pandq}
\p & = & g^2 p^4 - E p^2 + 2Np - P^2 +\alpha ~,\nonumber\\
\q & = & g^2 q^4 - E q^2 + 2Mq - Q^2 +\alpha~.
\eea
Here we have assumed a negative cosmological constant $\Lambda = -3g^2$, as in (\ref{EOM}), 
and $E, \alpha, M, N, P$ and $Q$ are arbitrary constants. The gauge field is
\bea
A &=& \frac{p P +q Q }{p^2 - q^2} \dd \tau + pq \frac{q P + p Q }{p^2 - q^2} \dd \sigma~,
\eea
which thus depends only on the parameters $P$ and $Q$. Moreover, one easily checks that 
when $P=\pm Q$ the gauge field $A$ has self-dual/anti-self-dual curvature $F=\diff A$ 
(depending on the choice of orientation), and that the metric (\ref{PDmetric}) is Einstein.

In \cite{AlonsoAlberca:2000cs} the authors studied which of the Plebanski-Demianski solutions 
above are supersymmetric solutions to the $d=4$, $\mathcal{N}=2$ gauged supergravity described in section 
\ref{gaugedsec}; that is, which admit non-trivial solutions to the Killing spinor equation (\ref{KSE}). 
This leads to the following BPS equations for the parameters:
\bea\label{BPS}
NQ + MP & = & 0~,\nonumber\\
 \left[N^2-M^2 - E(P^2-Q^2)\right]^2  & =  & 4g^2 \alpha (P^2-Q^2)^2~.
\eea
These arise from the BPS equations as presented in \cite{AlonsoAlberca:2000cs}, on making the Euclidean change of variables 
described in the footnote above. 

For applications to the AdS/CFT correspondence one is interested in solutions 
which have an asymptotic conformal boundary. It is then natural to assume
that either $p$ or $q$ is the radial variable near this boundary, and without loss of generality 
we  take this to be $q$. As $q\rightarrow\pm \infty$ the metric (\ref{PDmetric}) tends to
\bea
g^2\diff s^2_4 &=& \frac{\dd q^2}{q^2} + q^2 \dd s^2_{3}~,
\eea
where the corrections are $O(1/q^2)$ relative to this metric, and the boundary three-metric is defined as 
\bea\label{conformalmetric}
\frac{1}{g^2}\dd s^2_3 & = & -\frac{\dd p^2}{\p}  - \p \dd\sigma^2 + g^2 (\dd\tau + p^2 \dd\sigma)^2~.
\eea
In principle one could now carry out a systematic analysis of which solutions to the BPS equations 
(\ref{BPS}) lead to a compact smooth boundary three-manifold of the form (\ref{conformalmetric}), with 
moreover a smooth interior metric (\ref{PDmetric}).\footnote{It is also important to 
ensure that the field strength $F$ is everywhere non-singular.} However, motivated by the 
field theory analysis on the $U(1)^2$-squashed sphere in \cite{Hama:2011ea}, we will 
content ourselves here by looking for a solution where the boundary three-metric (\ref{conformalmetric})
 takes the form (\ref{HHLintro}). We intend to return to the more general problem in 
future work.

We begin by noting that the polynomial $\p$ in (\ref{pandq}) may be written
\bea
\p &=& g^2(p^2-p_1^2)(p^2-p_2^2) + 2Np~.
\eea
This hence reduces to a simple quadratic in $p^2$ when $N=0$. Assuming the latter, we may then introduce 
coordinates
\bea
\frac{p^2 - p_1^2 }{p_2^2 - p_1^2 } & = & \cos^2 \theta ~,\qquad \frac{p^2_2 - p^2 }{p_2^2 - p_1^2 } \ = \ \sin^2 \theta ~,\nonumber\\
\sigma &=& \frac{1}{g^2(p_2^2-p_1^2)}\left(\frac{1}{p_1}\varphi_1 - \frac{1}{p_2}\varphi_2\right)~,\nonumber\\
\tau &=& \frac{1}{g^2(p_2^2-p_1^2)}\left(-p_1\varphi_1 +p_2\varphi_2\right)~,
\eea
to obtain the boundary metric
\bea
\dd s^2_3 & = & \frac{\dd \theta^2}{ p_2^2 \cos^2 \theta + p_1^2\sin^2\theta} + \frac{1}{p_1^2}\cos^2\theta
\dd \varphi_1^2 +  \frac{1}{p_2^2}\sin^2\theta \dd \varphi_2^2~.
\label{nicemetric}
\eea
Multiplying by $p_1^2$ and identifying $p_2/p_1=\s$ then precisely leads to our boundary metric (\ref{squashedS3}). 
Notice that all we have assumed to obtain this result is $N=0$.

Of course, we must then find a smooth filling of this boundary metric. Our four-dimensional metric and gauge field (\ref{soln}) 
arise from the solution 
\bea\label{BPSregular}
M &=& 0~, \qquad E^2 \ =\ 4g^2\alpha~, \qquad P \ = \ - Q
\eea
of the BPS equations (\ref{BPS}). The coordinates in (\ref{soln}) are obtained by making the additional rescalings
\bea
p &=& p_1x~, \qquad q \ = \ p_1 y~, \qquad \tau \ = \ \frac{1}{p_1}\Psi~, \qquad \sigma \ = \ \frac{1}{p_1^3}\Phi~.
\eea
It is not difficult to see that (\ref{soln}), or equivalently (\ref{BPSregular}), is the \emph{only} regular 
solution of the BPS equations (\ref{BPS}), although this involves analysing a number of subcases 
and we omit the details. Of course, in any case in principle one should show  that 
(\ref{soln}) is the unique regular solution of the Einstein-Maxwell equations with appropriate boundary conditions, not just the unique solution within the supersymmetric Plebanski-Demianski class. 
This uniqueness question has been addressed in the mathematics literature for Einstein 
metrics -- see, for example, \cite{Anderson:2006ax, Anderson, AndersonHerzlich} --  but we are not aware of any detailed work on the problem in Einstein-Maxwell theory.

\section{Supergravity Killing spinor}

In this appendix we give some further details of the Killing spinor computation in section \ref{susysec}. 
It is straightforward to substitute the metric and gauge field (\ref{soln}) into the Killing spinor equation (\ref{KSE}), 
using the orthonormal frame (\ref{frame}) and explicit basis of $\mathrm{Cliff}(4,0)$ given in (\ref{gammas}). 
In particular, one extracts the following $y$ and $x$ components of the Killing spinor equation:
\bea
\partial_y  \epsilon^- + \frac{f_2}{f_1}\frac{1}{2(y-x)} \ii \sigma_3 \epsilon^- - \frac{g f_2}{2} \ii \mathbb{I}_2 \epsilon^+ & =  & 0~,\label{diffy}\\
\partial_y \epsilon^+ +\frac{f_2}{f_1}\frac{1}{2(y+x)} \ii \sigma_3 \epsilon^+ + \frac{gf_2}{2} \left( \ii \mathbb{I}_2  + w \ii \sigma_2 \right) \epsilon^- & = & 0 ~, \label{diffy2}\\
\partial_x \epsilon^- +\frac{f_1}{f_2}\frac{1}{2(y-x)} \ii \sigma_3 \epsilon^-   + \frac{g f_1}{2} \sigma_3 \epsilon^+ & = & 0  ~,\label{diffx} \\
\partial_x \epsilon^+ - \frac{f_1}{f_2}\frac{1}{2(y+x)} \ii \sigma_3 \epsilon^+   + \frac{gf_1}{2} \left(  \sigma_3 + w \ii \sigma_1 \right) \epsilon^- & = & 0 \label{diffx2}~.
\eea
Here we have defined the function
\bea
w(x,y) &\equiv & \frac{\s^2-1}{(y+x)^2}~,
\eea
so that the gauge field curvature is 
\bea
F &=& \frac{gw}{2} (e^{13} + e^{24})
\eea
in the frame (\ref{frame}). Using the algebraic relation (\ref{algebraic}), which recall follows from the 
integrability condition for the Killing spinor equation, we may eliminate $\epsilon^+$ from (\ref{diffy}) and (\ref{diffx}), 
leading to
\bea\label{dy}
\left[ \partial_y - \frac{1}{2(y+x)} + \frac{f_2}{f_1} \frac{y}{y^2-x^2} \ii \sigma_3 \right] \epsilon^-  & = & 0~, \\\label{dx}
\left[ \partial_x - \frac{1}{2(y+x)} + \frac{f_1}{f_2} \frac{x}{y^2-x^2} \ii \sigma_3 \right] \epsilon^-  &= & 0~.
\eea
Since the Pauli matrix $\sigma_3$ is diagonal, equations (\ref{dy}), (\ref{dx}) lead to decoupled 
equations for the two components of $\epsilon^-$. We thus write
\bea
\epsilon^- &=& \left(\begin{array}{c}\epsilon^-_+\\ \epsilon^-_-\end{array}\right)~,
\eea
so that (\ref{dy}), (\ref{dx}) are equivalent to the four equations
\bea\label{ODEy}
\partial_y \epsilon^-_\pm + Y_\pm(x,y)\epsilon^-_\pm&=&0~,\\\label{ODEx}
\partial_x \epsilon^-_\pm+X_\pm(x,y)\epsilon^-_\pm&=&0~,
\eea
where we have defined 
\bea
Y_\pm(x,y)&\equiv & -\frac{1}{2(y+x)} \pm \frac{f_2}{f_1} \frac{\ii y}{y^2-x^2} ~,\\
X_\pm(x,y) &\equiv & -\frac{1}{2(y+x)} \pm \frac{f_1}{f_2} \frac{\ii x}{y^2-x^2} ~.
\eea
The integrability condition for (\ref{ODEy}), (\ref{ODEx}) is $\partial_x Y_\pm(x,y)=\partial_y X_\pm(x,y)$, which 
is easily verified to hold. These are then  first order linear homogeneous differential equations, which may be integrated 
to give
\bea\label{barry}
\epsilon^-_\pm &=& c_\pm \sqrt{y+x}\left(\frac{\sqrt{(\s^2-x^2)(y^2-1)}\mp\ii \sqrt{(x^2-1)(y^2-\s^2)}}{\sqrt{(\s^2-x^2)(y^2-1)}\pm\ii \sqrt{(x^2-1)(y^2-\s^2)}}\right)^{1/2}~,
\eea
where $c_\pm$ are integration constants ({\it a priori} depending on the angular coordinates $\Psi$ and $\Phi$). 

One can now substitute the solutions (\ref{barry}) into the remaining differential equations (\ref{diffy2}), (\ref{diffx2}), 
which one finds are satisfied if and only if
\bea
c_- &=& \ii c_+~,
\eea
which leads to the form of the Killing spinor given in (\ref{epplus}). Finally, from the 
$\Psi$ and $\Phi$ components of the Killing spinor equation it is reasonably simple to extract $\partial_\Psi c_+=\partial_\Phi c_+=0$, 
so that the spinor $\epsilon$ is independent of 
$\Psi$ and $\Phi$. A somewhat more lengthy calculation then confirms that the remaining components 
of the Killing spinor equation are all satisfied.

\label{Killappendix}

\section{One-loop vector multiplet contribution}\label{vectorapp}

In this appendix, for completeness we explain how to show the equality between the middle and last lines of equation (\ref{detvectornice}). 
We begin with a trick similar to that used in \cite{Kapustin:2009kz}: the eigenvalues of a matrix in the adjoint 
representation come in positive-negative pairs, so that (\ref{detvectornice}) is even in $\sigma$.  
This implies that one can equivalently sum over only the \emph{positive} roots in the middle line of (\ref{detvectornice}), 
while at the same time multiplying the right-hand side by itself with $\sigma\mapsto -\sigma$, to obtain the same result.
This leads to the equality
\bea
Z_{\mathrm{one-loop}}^{\mathrm{vector}}(\sigma) & = & \prod_{\mathrm{positive\, roots}} \frac{1}{\alpha(\sigma)^2}\prod_{m,n\geq 0} \Bigg[
\frac{m+nb^2 + \ii\alpha(\sigma)}{m+1+(n+1)b^2 - \ii\alpha(\sigma)}\cdot \nn\\
&& \frac{m+nb^2 - \ii\alpha(\sigma)}{m+1+(n+1)b^2 + \ii\alpha(\sigma)}\Bigg]~.
\eea
Next notice that (formally) all the numerator terms cancel against denominator terms in the product over all $m,n\geq 0$, 
\emph{except} for the numerator contributions of $\{m=0, n=0\}$, $\{m=0, n\geq 1\}$ and $\{m\geq 1, n=0\}$, which are left uncancelled. 
The first of these cancels the $\alpha(\sigma)^2$ prefactor, and we  immediately reduce to 
\bea
Z_{\mathrm{one-loop}}^{\mathrm{vector}}(\sigma) & = & \prod_{\mathrm{positive\, roots}}\, \prod_{n\geq 1} \, (n^2+\alpha(\sigma)^2)(n^2b^4+\alpha(\sigma)^2)~.
\eea
The above manipulations are somewhat formal, as this is clearly divergent. However, we may write
\bea
Z_{\mathrm{one-loop}}^{\mathrm{vector}}(\sigma) & = & \prod_{\mathrm{positive\, roots}}\left(\prod_{n\geq 1} b^4n^4\right)\prod_{n\geq 1}\left(
1+\frac{\alpha(\sigma)^2}{n^2}\right)\left(
1+\frac{\alpha(\sigma)^2}{b^4n^2}\right)~,
\eea
and then use the product formula for $\sinh (\pi z)$:
\bea
\sinh (\pi z)  & = & \pi z \prod_{n=1}^\infty \left(1+ \frac{z^2}{n^2}\right)
\eea
for the last product. Using the zeta function regularization,  the divergent prefactor is (for $b\neq 0$)
\bea
\prod_{n\geq 1} b^4n^4 & \stackrel{\mathrm{zeta}\, \mathrm{reg}}{=} & \frac{(2\pi)^2}{b^2}~. 
\eea
Putting everything together then gives the last line of (\ref{detvectornice}). Notice we have corrected a factor of $\pi^2$ compared to the corresponding
formula in the original reference \cite{Kapustin:2009kz}.

Finally, we note that the denominator in the last line  of (\ref{detvectornice}) in fact cancels against the Vandermonde determinant when reducing the
integral from the Lie algebra to its Cartan subalgebra. More precisely and specifically, the Haar measure for $U(N)$ is
\bea
\diff\mu &=& \prod_{i=1}^N \diff \sigma_i \, \Delta(\sigma)^2
\eea
where $\sigma_i$ denote the eigenvalues of $\sigma$ (so $\sigma_i=\frac{\lambda_i}{2\pi}$) and $\Delta(\sigma)$ is the Vandermonde determinant
\bea
\Delta(\sigma) &=& \prod_{i<j}\, (\sigma_i-\sigma_j) \ = \ \prod_{\mathrm{positive\, roots}} \alpha_{ij}(\sigma)~.
\eea

\end{document}